\documentclass[manuscript,screen,natbib,nonacm]{acmart}

\AtBeginDocument{%
  }

\setcopyright{acmlicensed}
\copyrightyear{2018}
\acmYear{2018}
\acmDOI{XXXXXXX.XXXXXXX}

\acmISBN{978-1-4503-XXXX-X/18/06}

\usepackage[T1]{fontenc}
\usepackage[utf8]{inputenc}
\usepackage[american]{babel}
\usepackage{csquotes}
\usepackage{microtype}
\usepackage{verbatim}

\usepackage{paralist}
\setdefaultenum{(1)}{(a)}{(i)}{A.}
\usepackage{enumitem}
\usepackage{url}
\usepackage{flushend}
\usepackage{etoolbox}
\usepackage{fix-cm}
\usepackage{textpos}
\usepackage{mfirstuc}
\usepackage{titlecaps}
\usepackage[datesep=.,style=ddmmyyyy]{datetime2}
\usepackage[colorinlistoftodos,prependcaption,textsize=tiny]{todonotes}

\usepackage{amsmath}
\usepackage{bm}
\usepackage{bbm}
\usepackage{bbold}
\usepackage{relsize}
\usepackage{siunitx}
\usepackage[super]{nth}
\usepackage{braket}
\usepackage{amsthm}
\usepackage{quantikz}

\usepackage{booktabs}
\usepackage{multirow}
\usepackage{colortbl}
\usepackage{makecell}
\usepackage{rotating}
\usepackage{threeparttable}
\usepackage{array}
\usepackage{siunitx}
\usepackage{longtable}

\usepackage{algorithm}
\usepackage{algpseudocode}
\usepackage{enumitem}
\usepackage{listingsutf8}

\usepackage{xparse}
\usepackage{xspace}

\usepackage[xindy,acronym]{glossaries}
\glsdisablehyper

\newcommand{\addgate}{\textsc{Add}}
\newcommand{\replacegate}{\textsc{Replace}}
\newcommand{\removegate}{\textsc{Remove}}

\usepackage{natbib}

\usepackage{hyperref}
\usepackage[all]{hypcap}

\usepackage[capitalise]{cleveref}
\usepackage{xr}

\newacronym{sut}{SUT}{system under test}
\newacronym{qst}{QST}{Quantum Software Testing}
\newacronym{qft}{QFT}{Quantum Fourier Transform}
\newacronym{ps}{PS}{program specification}
\newacronym{sqqpt}{SQQPT}{Single Qubit Quantum Process Tomography}
\newacronym{qpe}{QPE}{Quantum Phase Estimation}
\newacronym{qpt}{QPT}{Quantum Process Tomography}
\newacronym{fmem}{FMEM}{Fault Matrix Estimation Mapping}
\newacronym{mwu}{MWU}{Mann-Whitney U}
\newacronym{iqr}{IQR}{Interquartile Range}
\newacronym{q1}{Q1}{Quartile 1}
\newacronym{q3}{Q3}{Quartile 3}
\newacronym{se}{SE}{Standard Error}
\newacronym{ci}{CI}{Confidence Interval}
\newacronym{sdk}{SDK}{Software Development Kit}
\newacronym{nisq}{NISQ}{Noisy Intermediate-Scale Quantum}
\newacronym{hhl}{HHL}{Harrow-Hassidim-Lloyd}
\newacronym{qaoa}{QAOA}{Quantum Approximate Optimization Algorithm}
\newacronym{vqe}{VQE}{Variational Quantum Eigensolver}
\newacronym{pp}{pp}{percentage point}

\newtheorem{definition}{Definition}

\DeclareMathOperator{\Tr}{Tr}

\definecolor{custombordergreen}{RGB}{37,116,59}  
\definecolor{customcontentgreen}{RGB}{148,180,162} 

\newcommand{\summarybox}[2]{%
  \noindent
  \fboxsep=7pt 
  \fcolorbox{custombordergreen}{customcontentgreen}{%
    \parbox{0.95\textwidth}{%
      \textbf{#1:} #2%
    }%
  }%
}

\begin{document}

\title{Bloch Vector Assertions for Debugging Quantum Programs}


\author{Noah H. Oldfield}
\orcid{0000-0002-9059-0694}
\affiliation{%
  \institution{Simula Research Laboratory and University of Oslo}
  \city{Oslo}
  \country{Norway}}
\email{noah@simula.no}

\author{Christoph Laaber}
\orcid{0000-0001-6817-331X}
\affiliation{%
  \institution{Simula Research Laboratory}
  \city{Oslo}
  \country{Norway}}
\email{laaber@simula.no}

\author{Shaukat Ali}
\orcid{0000-0002-9979-3519}
\affiliation{%
  \institution{Simula Research Laboratory}
  \city{Oslo}
  \country{Norway}}
\affiliation{%
  \institution{Oslo Metropolitan University}
  \city{Oslo}
  \country{Norway}}
\email{shaukat@simula.no}



\begin{abstract}
Quantum programs must be reliable to ensure trustworthy results, yet debugging them is notoriously challenging due to quantum-specific faults like gate misimplementations and hardware noise, as well as their inherently probabilistic nature.
Assertion-based debugging provides a promising solution by enabling localized correctness checks during execution. 
However, current approaches face challenges including manual assertion generation, reliance on mid-circuit-measurements, and poor scalability.
In this paper, we present Bloq, a scalable, automated fault localization approach introducing Bloch-vector-based assertions utilizing expectation value measurements of Pauli operators, enabling low-overhead fault localization without mid-circuit measurements.
In addition, we introduce AutoBloq, a component of Bloq for automatically generating assertion schemes from quantum algorithms.
An experimental evaluation over \num{684432} programs using two algorithms (\gls{qft} and Grover) shows that Bloq consistently outperforms the state-of-the-art approach Proq, notably as circuit depth and noise increase.
For Grover, Bloq achieves a mean F1 score across all experimental instances of \num{0.74} versus \num{0.38} for Proq under ideal conditions, and maintains performance under noise (\num{0.43} versus \num{0.06}).
Bloq also reduces Proq's runtime by a factor of 5 and circuit depth overhead by a factor of 23.
These results underline Bloq’s potential to make assertion-based debugging scalable and effective for near-term quantum devices.
\end{abstract}



\maketitle

\newcommand{\HLIGHT}[1]{``#1''}

\section{Introduction}
\label{sec:introduction}

Quantum computing offers the potential for vast performance improvements in domains such as optimization, quantum chemistry, and cryptography~\citep{feynman2018simulating,QCQIBook,Egger_2020,moll2018quantum,GoogleAILearning,Shor_1997}. 
However, quantum programs are highly error-prone due to the fragile nature of quantum states and the inherent noise in current quantum hardware~\citep{QCQIBook}, in addition to quantum software faults, such as those caused by incorrect gate operations in the quantum algorithm~\citep{BugsinQCEmpirical,campos2021qbugs,aoun2022bug}. 
Ensuring the correctness of quantum programs is therefore a fundamental challenge in the emerging field of quantum software engineering~\citep{ShaukatTaoQST}.

Assertion-based runtime testing and debugging, where assertions are inserted into a quantum program to validate expected conditions during execution, has emerged as a promising approach for fault localization in quantum software~\citep{Proq,Stat,QECA2020,LiExploitingAssertions,willeAssertionRefining}. 
However, existing techniques for runtime assertion testing face four significant challenges:

\begin{enumerate}
    \item \textbf{Assertion Depth Overhead:} 
    The Proq approach of \citet{Proq} frequently requires a complete uncomputation of the circuit, which increases the depth of the program with each assertion.
    \item  \textbf{Compatibility with \gls{nisq} Systems}
    First, the inherent depth overhead in projective measurements is particularly problematic for current \gls{nisq} systems due to the accumulation of errors with circuit depth.
    Second, projective measurements rely on mid-circuit measurements, which are particularly demanding for quantum hardware because they require interrupting circuit execution to perform real-time measurements, increasing both complexity and execution cost~\citep{ChenMidCircuit}.
    \item \textbf{Lack of Automation:} 
    While theoretically effective for debugging, identifying appropriate assertions is challenging due to the reliance on deep understanding of the algorithm's mathematical properties, ad hoc inspection, or experimental methods~\citep{willeAssertionRefining}.
    \item \textbf{Need for Empirical Evaluations:} Existing assertion-based approaches have only been evaluated in a limited number of studies. 
    Proq~\citep{Proq} and Stat~\citep{Stat} present case studies, while QECA~\citep{QECA2020} provides an experimental evaluation on small programs of 2 and 3 qubits. 
    This highlights the need for more comprehensive empirical evaluations to assess the precision, scalability, and runtime cost of assertion-based debugging.
\end{enumerate}

To address these challenges, we propose Bloq, a scalable and automated fault localization approach for quantum programs based on Bloch vector assertions. 
Bloq addresses challenge (1) by leveraging expectation value measurements of single-qubit Pauli operators, avoiding the need for depth-increasing gates or mid-circuit measurements. 
To address (2), Bloq relies solely on Pauli measurements, which are readily available on current \gls{nisq} hardware.
Challenge (3) is addressed through AutoBloq, a procedure for automatically generating assertion schemes from quantum program specifications. 
Finally, to address (4), we provide an extensive empirical evaluation of Bloq on both ideal and noisy simulators, comparing its performance against the state-of-the-art projective measurement-based approach Proq~\cite{Proq}, across fault localization effectiveness, scalability, and runtime cost.

Our evaluation assesses Bloq’s effectiveness across qubit count, circuit depth, fault type, and fault location, as well as its runtime and circuit depth overhead, by comparing it against Proq, a projective-measurement-based approach used as our baseline, on two representative quantum programs: Grover’s algorithm and \gls{qft}.
In total, we evaluate \num{684432} quantum program instances, covering all combinations of circuit inputs, fault location, fault types and fault categories.
We find that Bloq consistently outperforms Proq for Grover, achieving an overall average F1 score of \num{0.74} compared to \num{0.38} with the ideal backend, and remaining significantly more effective under noise with \num{0.43} versus \num{0.06}. 
Bloq’s effectiveness increases with qubit count, reaching an F1 score of \num{0.85} at 6 qubits, while Proq’s performance declines to \num{0.15}. 
For \gls{qft}, both approaches perform similarly in smaller circuits, but Bloq begins to outperform Proq in higher-depth circuits, noisier programs, achieving \num{0.34} at 10 qubits compared to \num{0.32}. 
In addition to improved fault localization, Bloq significantly reduces runtime and circuit depth overhead: for Grover, Bloq’s circuits are nearly 23$\times$ lower-depth on average (\num{229.85} versus \num{5287.51}) and execute up to 5$\times$ faster under noise (\SI{28.91}{\second} vs. \SI{143.61}{\second}). 
Finally, Bloq enables automated assertion generation via AutoBloq, addressing a key usability limitation in prior assertion-based testing approaches.

The contributions of this paper are as follows:

\begin{enumerate}
    \item A definition of Bloch vector-based runtime assertions (Bloq) for fault localization in quantum programs, leveraging expectation value measurements of single-qubit Pauli operators to enable low-overhead testing without depth-increasing gates or mid-circuit measurements.
    
    \item AutoBloq, a scalable, program-independent, and automated procedure for generating assertion schemes from quantum program specifications, enabling the production of on-demand assertion values for any test input.
    
    \item An extensive experimental evaluation framework of Bloq's fault localization effectiveness, comparing against the state-of-the-art projective measurement-based approach Proq~\cite{Proq}. 
    The evaluation covers two representative quantum programs, \gls{qft} and Grover's algorithm, under both ideal and noisy conditions using IBM's AerSimulator backend with the noise model of ibm\_sherbrooke~\cite{ibm_quantum_2021}.
\end{enumerate}

A replication package containing all code, data, and analysis scripts used in this paper is available online~\cite{Bloq_replication}.

\section{Background}

In this section, we introduce the necessary background information on quantum computing.

\subsection{Density Matrix Representation}

All states that can be represented by the common state vector representation are called \textit{pure states} \cite{QCQIBook}.
These are states where we have complete information about the quantum system.
This is not always the case, however, if there is noise in the quantum computer or if the algorithm simply contains steps where the single qubit states are mixed due to entanglement.
Thus, to represent general quantum states in quantum computing, the \textit{density matrix} representation is a common choice.
Unlike the state vector representation, the density matrices can represent both \textit{pure} and \textit{mixed} quantum states:

\begin{equation}
    \rho = \sum_j \ket{\psi_j}\bra{\psi_j}
\end{equation}

Pure states are those states that we have perfect information about, which can be expressed also as state vectors.
While mixed states cannot be written as state vectors, but are mixtures of arbitrary pure states \cite{QCQIBook}.

To determine experimental outcomes of our quantum program, we define expectation values of observables by:

\begin{equation}
    \braket{A} = Tr(A\rho)
\label{eq:expectation_value_of_observable}
\end{equation}

In \cref{eq:expectation_value_of_observable}, $\braket{A}$ is the expectation value of the observable $A$, which is a linear operator over the state Hilbert space.

Our main results will heavily rely on the following property of density matrices.
Any single qubit state can be written in the following form in the density matrix representation:

\begin{equation}
    \rho = \frac{1}{2}\Big( \mathbbm{1} + \braket{X}X + \braket{Y}Y + \braket{Z}Z\Big)
    \label{eq:density_matrix_representation_single_qubit}
\end{equation}

Where $\mathbbm{1}$ is the identity matrix and $\braket{X}, \braket{Y}, \braket{Z}$ are the Pauli matrices:

\begin{align}
X &= \begin{pmatrix}
        0 & 1\\
        1 & 0
    \end{pmatrix} &     Y &= \begin{pmatrix}
        0 & -i \\
        i & 0
    \end{pmatrix}  &   Z &= \begin{pmatrix}
        1 &  0 \\
        0 & -1
    \end{pmatrix}
\end{align}

The density matrix is a strictly positive matrix and satisfies normality, $\Tr \rho = 1$. 
In addition, if $\rho$ is a pure state, then $\Tr \rho^2 =1$, and if it is a mixed state, then $\Tr \rho^2 <1$.
Thus, $\rho$ satisfies $\Tr \rho^2 \leq 1$.
Furthermore, we employ a vector called the \textit{Bloch} vector, which compacts and simplifies the calculations and results in the later states of our paper. 
The single qubit state can be parameterized by the Pauli expectation values over the Bloch sphere such that the vector:
\begin{equation}
    \Vec{R} = \Big( \braket{X}, \braket{Y}, \braket{Z}\Big)
    \label{eq:bloch_vector}
\end{equation}

In \cref{eq:bloch_vector}, we call $\Vec{R}$ the \textit{Bloch vector} and the Pauli expectation values specify the Bloch sphere coordinates of $\rho$ \cite{QCQIBook}.

\subsection{Quantum State Fidelity}
\label{sec:background_fidelity}

Quantum state fidelity is a widely used metric for quantifying the similarity between two quantum states~\citep{QCQIBook}. 
Given two density matrices $\rho$ and $\sigma$, the fidelity between them is defined as:

\begin{equation}
    F(\rho, \sigma) = \left( \Tr \sqrt{ \sqrt{\rho} \sigma \sqrt{\rho} } \right)^2.
    \label{eq:state_fidelity}
\end{equation}

Fidelity takes values in the range $[0,1]$, where $F=1$ indicates that $\rho$ and $\sigma$ are identical, and $F=0$ indicates that the states are perfectly distinguishable.

In our evaluation, we use fidelity to measure the similarity between the expected density matrix and the measured density matrix at assertion points in the quantum circuit under test.

\subsection{Quantum Algorithms}
\label{sec:background_quantum_algorithms}

Inspired from a common way of writing down a quantum algorithm~\citep{QCQIBook,GroversAlgo}, we observe that a quantum algorithm follows a typical structure which we can write down as the following:

\paragraph{\textbf{Quantum Algorithm:}}
\label{sec:background_algorithm_spec}
\begin{enumerate}
    \item Initial Stage
    \item Create Superposition Stage
    \item iterative step
    \item Measurement Stage
\end{enumerate}

In the initial stage, we simply initialize the program to the single input state.
Next, in the create superposition stage, we apply Hadamard gates to create a superposition of the provided input state.
Then, in the iterative step, we perform iterations of a given operation on the superposition state.
Finally, we measure the state to obtain a single output state.
We use this algorithm structure in \cref{sec:approach} to define a program segment which we apply in our approach.

Examples of quantum algorithms which follow this structure are \gls{qft} and Grover's algorithm, which we briefly define here in the background, before we utilize them in our evaluation in \cref{sec:evaluation}.
This iterative structure, initialization, repeated transformation, and measurement, also appears in other quantum algorithms such as quantum phase estimation, amplitude amplification, and quantum walks~\cite{QCQIBook,Venegas_Andraca_2012_Quantum_Walk,GroversAlgo}.
In contrast, algorithms such as Deutsch-Jozsa and Simon’s are non-iterative and mainly of theoretical interest. 
Shor’s algorithm combines both patterns: while its quantum circuit is not repeated, it applies a modular sequence of unitaries and is typically rerun to extract the period~\cite{Shor_1997}.

\paragraph{\textbf{Quantum Fourier Transform}}

\Gls{qft} is a fundamental quantum algorithm that performs a discrete Fourier transform on the amplitudes of a quantum state. 
It is widely regarded as a key building block for many quantum algorithms, most notably Shor's algorithm for factoring integers and finding discrete logarithms. \Gls{qft} operates as a modular component within these algorithms, enabling efficient computation of periodicity and phase relationships that are crucial for their functionality.

The \gls{qft} circuit consists of a sequence of controlled rotation gates and Hadamard gates applied to the input state. 
It takes an input state \(\ket{j} = \ket{j_0j_1 \cdots j_{n-1}}\) and maps it to the corresponding output state vector using a transformation described in conventional quantum computing literature~\citep{QCQIBook}:

\begin{equation}
    \ket{j_0 j_1 \cdots j_{n-1}} = \frac{1}{\sqrt{2^n}}\Big( \ket{0} + e^{2\pi i 0.j_{n-1}}\ket{1}\Big)\Big( \ket{0} + e^{2\pi i 0.j_{n-2}j_{n-1}}\ket{1}\Big)\cdots \Big( \ket{0} + e^{2\pi i 0.j_0 j_1 \cdots j_{n-1}}\ket{1}\Big).
    \label{eq:qft_transformation}
\end{equation}

Here, the notation \(j = j_0j_1\cdots j_{n-1}\) represents the binary decomposition of the decimal number \(j\). 
For an \(n\)-qubit input state, \(j\) is a decimal integer ranging from \(0\) to \(2^n - 1\), and its binary representation is given by \(j_0j_1\cdots j_{n-1}\), where \(j_0, j_1, \dots, j_{n-1}\) are the bits of \(j\) in decreasing significance. For example:
\[
j = 6 \implies j_0j_1j_2 = 110 \quad (\text{binary representation of } 6).
\]
The phase factors \(0.j_{n-1}, 0.j_{n-2}j_{n-1}, \dots, 0.j_0j_1\cdots j_{n-1}\) are fractional binary numbers, representing the binary digits \(j_{n-1}, j_{n-2}, \dots, j_0\) as terms in a binary fraction. 
For instance:
\[
0.j_{n-1} = \frac{j_{n-1}}{2}, \quad 0.j_{n-1}j_n = \frac{j_{n-2}}{2} + \frac{j_{n-1}}{4}, \quad \text{and so on.}
\]
These phase terms encode the periodicity and interference patterns that \gls{qft} exploits for its applications in quantum algorithms.
The iterative step of the \gls{qft} circuit consists of applying a Hadamard gate to a qubit, followed by a sequence of controlled phase rotations entangling it with less significant qubits, resulting in the transformed qubit states in \cref{eq:qft_transformation}.

\paragraph{\textbf{Grover's Algorithm}}

Grover's algorithm is a quantum search algorithm that provides a quadratic speedup over classical search algorithms for unstructured databases~\citep{GroversAlgo}. 
It finds a marked item within an unsorted dataset of \(N = 2^n\) items using \(O(\sqrt{N})\) queries, compared to the \(O(N)\) complexity of classical algorithms. 
Grover's algorithm is widely regarded as one of the most practical examples of quantum advantage due to its general applicability in search problems, optimization, and constraint satisfaction tasks.

The algorithm applies the \emph{Grover operator} \(G\) in the iterative step, amplifying the amplitude of the marked state while reducing the amplitude of unmarked states.
The operator is defined as:
\begin{equation}
    G = (2\ket{\psi}\bra{\psi} - I)O,
    \label{eq:grover_operator}
\end{equation}
where \(\ket{\psi}\) is the equal superposition state, \(I\) is the identity operator, and \(O\) is the oracle operator that marks the target state by flipping its phase.

Grover’s algorithm consists of the following steps:
\begin{enumerate}
    \item \emph{Initialization}: The system is initialized into a uniform superposition state:
   \[
   \ket{\psi} = \frac{1}{\sqrt{N}} \sum_{x=0}^{N-1} \ket{x}.
   \]
    \item \emph{Oracle Application}: The oracle \(O\) flips the phase of the target state \(\ket{w}\), where \(w\) is the marked item in the search space:
   \[
   O\ket{x} = 
   \begin{cases} 
   -\ket{x}, & \text{if } x = w, \\
   \ket{x}, & \text{otherwise.}
   \end{cases}
   \]
    \item \emph{Amplitude Amplification}: The Grover operator \(G\) increases the probability amplitude of the marked state while reducing that of the unmarked states.
\end{enumerate}

The algorithm repeats the application of \(G\) approximately \(O(\sqrt{N})\) times, after which a measurement collapses the state to the target state \(\ket{w}\) with high probability.

Grover's algorithm represents a distinct class of quantum programs compared to algorithms like \gls{qft}. Unlike \gls{qft}, which operates on exponentially large output distributions, Grover produces a single marked output state, making it ideal for demonstrating the versatility of Bloq assertions across different algorithmic classes.

\subsection{Projection-Based Assertion Scheme}
\label{sec:background_projection_schemes}

Projection-based runtime assertions are a class of quantum test oracles that validate program correctness by briefly reversing computation to inspect intermediate qubit states. 
These assertions follow a three-stage implementation process:

\begin{enumerate}
    \item \textbf{UNCOMPUTE}: An inverse operation is applied to undo the quantum transformation on a target qubit.
    \item \textbf{MEASUREMENT}: A mid-circuit measurement is performed on the uncomputed qubit to observe whether it returns to its expected basis state.
    \item \textbf{RECOMPUTE}: The original operation is reapplied to restore the program’s execution state.
\end{enumerate}

This technique is most effective when the expected state of a qubit is known or controllable, such as in algorithms with structured inputs. 
For example, in \gls{qft}, applying the inverse of a known single-qubit transformation allows to verify whether a qubit returns to its original computational basis state.
The observed measurements are typically stored in dedicated classical registers and compared to the expected outcomes, enabling localized fault detection without requiring full output verification. 
In ideal, fault-free conditions, projection-based assertions are expected to yield the original input values with high probability.

\section{Approach}
\label{sec:approach}

This section outlines our Bloch vector-based fault localization approach (Bloq), in three subsections:
\begin{inparaenum}
    \item We provide an overview of our fault localization approach, and how we utilize Autobloq to generate Bloq assertions and how we apply them in our test engine.
    \item We define Bloch vector assertions (Bloq assertions) and describe how they are applied to localize faults within specific parts of the circuit, which we define in the following subsections as \emph{segments} of the quantum program.
    \item We define a test assessment for Bloq that applies quantum state fidelity to compare the measured state with the theoretical density matrix, and flags a test as failing if the deviation exceeds a defined threshold.
    \item Finally, we introduce AutoBloq, a method for deriving assertion schemes automatically using mathematical formulations. 
\end{inparaenum}

While \cref{fig:approach_overview_diagram} presents the execution flow of Bloq, including AutoBloq’s role in generating assertion schemes, we introduce Bloq assertions first to establish the conceptual foundations of Bloq, before describing their automation.

\subsection{Approach Overview}
\begin{figure}
    \centering
    \includegraphics[width=0.7\textwidth]{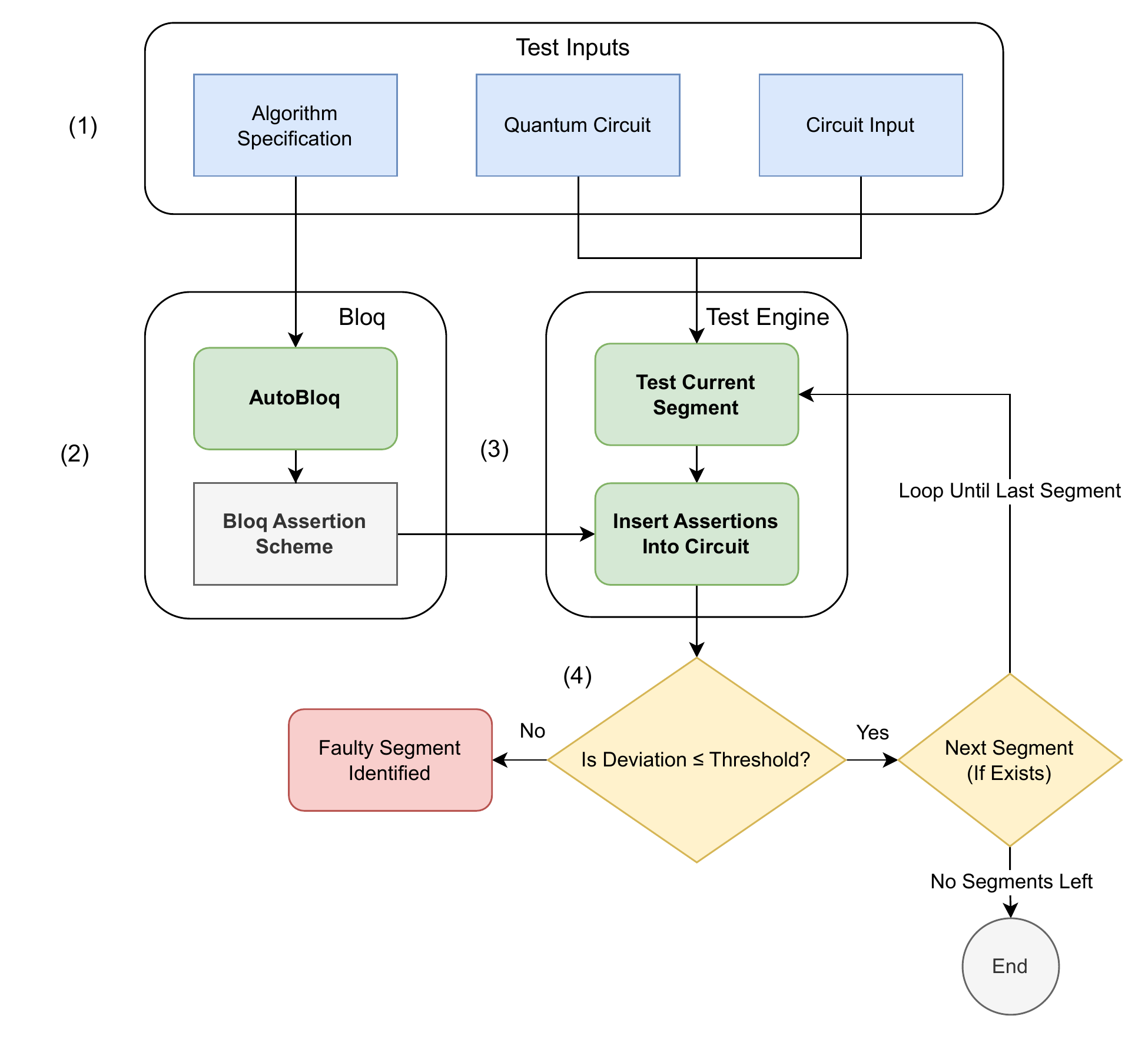}
    \caption{Overview of Bloq. The diagram depicts how test inputs are processed through AutoBloq to generate Bloq assertion schemes, which are applied by the test engine to identify faulty segments based on deviations from expected behavior.}
\label{fig:approach_overview_diagram}
\end{figure}
\Cref{fig:approach_overview_diagram} presents a high-level overview of our fault localization approach. 
The process begins with the \HLIGHT{Test Inputs} at (1), which include a \HLIGHT{Quantum Circuit}, an \HLIGHT{Algorithm Specification}, and a \HLIGHT{Circuit Input}. 
These inputs are passed to two independent components: the \HLIGHT{AutoBloq} tool, which generates a \HLIGHT{Bloq Assertion Scheme}, and the test execution process, which iterates over segments and qubits of the circuit to apply assertions.
In the first branch (2), labeled \HLIGHT{Bloq}, the \HLIGHT{Algorithm Specification} is passed to \HLIGHT{AutoBloq}, our method for generating an assertion scheme, referred to as the \HLIGHT{Bloq Assertion Scheme}. 
In the second branch at (3), labeled \HLIGHT{Test Engine}, we initialize the fault localization process by running the circuit up to the first segment and performing a Bloq assertion at that segment with the \HLIGHT{Test Current Segment} activity. 
Following this, we then utilize the \HLIGHT{Bloq Assertion Scheme}, received as input from component (2) and apply assertions to all qubits within the segment using the activity \HLIGHT{Insert Assertions Into Circuit}.
After the assertion is applied, the process reaches the decision point \HLIGHT{Is Deviation~$\leq$~Threshold?} at (4), where we perform our test assessment.
Here, we compute the deviation between the expected density matrix defined in the \HLIGHT{Bloq Assertion Scheme} and the measured density matrix. 
If any of the deviations for a Bloq assertion within the current segment are less than or equal to an assertion threshold.
We define this threshold as a percentage in the coming evaluation subsection, which accounts for sampling variability and backend noise.
In the case of a failing assertion at (4), a fault is detected, and the \HLIGHT{Faulty Segment Identified} activity outputs the test result, consisting of the index of the faulty segment and the qubit.
Thus, we localize the segment and the qubit where the assertion failed.
Otherwise, if all deviations remain within the threshold, the process continues to the next segment via the \HLIGHT{Next Segment (If Exists)} decision, which loops back to \HLIGHT{Test Current Segment} in the \HLIGHT{Test Engine} or exists the test to \HLIGHT{End} if all segments have been tested.

In the remainder of this section, we detail the two components introduced in this overview and illustrated in \cref{fig:approach_overview_diagram}: \HLIGHT{Bloq Assertions} and \HLIGHT{AutoBloq}.

\subsection{Bloch Assertions}
\label{sec:bloq_assertions_for_fault_loc}

\begin{figure*}[tbp]
    \centering
\begin{center}
\begin{quantikz}
\ket{j_0} & \qw & \gate[6]{ALG_0} \gategroup[6,steps=1,style={dashed,rounded
corners,fill=blue!20, inner
xsep=2pt},background,label style={label
position=below,anchor=north,yshift=-0.2cm}]{{\sc
segment 0}} & \qw & \gate[6]{ALG_1} \gategroup[6,steps=1,style={dashed,rounded
corners,fill=blue!20, inner
xsep=2pt},background,label style={label
position=below,anchor=north,yshift=-0.2cm}]{{\sc
segment 1}} & & \cdots & \qw & \gate[6]{ALG_k} \gategroup[6,steps=1,style={dashed,rounded
corners,fill=blue!20, inner
xsep=2pt},background,label style={label
position=below,anchor=north,yshift=-0.2cm}]{{\sc
segment $k$}}  & \qw & \qw &\cdots & \gate[6]{ALG_{\mathcal{K}-1}} \gategroup[6,steps=1,style={dashed,rounded
corners,fill=blue!20, inner
xsep=2pt},background,label style={label
position=below,anchor=north,yshift=-0.2cm}]{{\sc
segment $\mathcal{K}-1$}} \\
\ket{j_2} & & \qw & \qw & \qw & \qw & \cdots & \qw & \qw & \qw &  \qw &\cdots &  \\
\vdots \\
\ket{j_q} & & \qw & \qw & \qw & \qw & \cdots & \qw & \qw & \qw  &  \qw & \cdots & \\
\vdots \\
\ket{j_{n-1}} & & \qw & \qw & \qw & \qw & \cdots & \qw  & \qw & \slice{$\rho_{q, k}$} \qw  &  \qw &\cdots &  
\end{quantikz}
\end{center}
    \caption{Diagram of a quantum algorithm \( ALG_j \) with segments indexed by \( j = 0, 1, \cdots, \mathcal{K}-1 \). The dashed red line indicates where the density matrix \( \rho_{q,k} \) of qubit \( q \) is defined at segment \( k \).}
    \label{fig:diagram_algorithmic_structure}
\end{figure*}
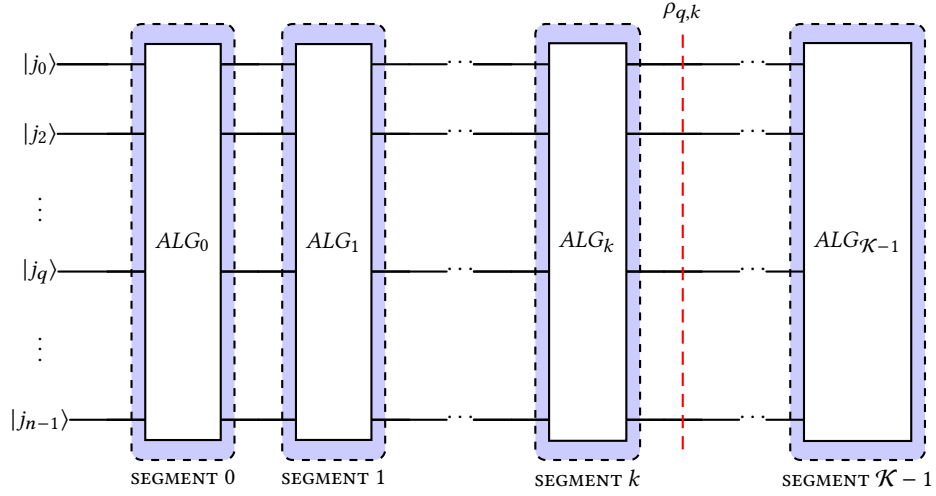

We design Bloq to perform runtime assertions at specific points in a quantum program, referred to as \textit{segments}, which demarcate distinct portions of the quantum circuit. 
While related approaches~\citep{BuggySegments,Proq} also utilize segments, they do not clearly define what constitutes a segment, often relying on ad hoc selection of assertion points.

To address this limitation, we first establish a clear and systematic definition of a program segment. 
This formalization streamlines the assertion generation process, enhancing both generalizability and automation for Bloq assertions.
We define a program segment as the portion of a quantum program resulting from a single application of the iterative step described in \cref{sec:background_algorithm_spec}.
Furthermore, in \cref{fig:diagram_algorithmic_structure}, we depict a single-qubit density matrix $\rho_{q,k}$, where $q$ denotes the qubit and $k$ denotes some arbitrary segment between \num{0} and $\mathcal{K}-1$.
A quantum program thus consists of a sequence of segments, denoted as $ALG_j$, where $j = 0, 1, \dots, \mathcal{K} - 1$, with $\mathcal{K}$ being the total number of segments.
Each segment $ALG_j$ represents the $j$th stage of the algorithm according to our defined algorithm specification in \cref{sec:background_algorithm_spec}.
For example, in \gls{qft}, the input state is transformed by applying a segment of rotation operators to each qubit, producing a final uniform superposition with quantized relative phases. 
In contrast, Grover's algorithm employs segments of the diffusion operator to perform an unstructured search on a specified oracle.

Runtime assertions for quantum programs have previously been defined by separate syntactic and semantic statements tailored specifically for projective measurements, along with suggested experimental techniques. 
Their approach applies a projective measurement operation to a segment of the program, followed by a validation step to verify the program's correctness~\citep{Proq}.
To increase the abstraction of runtime assertions beyond projective measurements, in addition to providing a better ease of use for software testers, we provide a single definition of runtime assertions to capture: syntax, quantum sampling, and predicate validation. 
This formulation is designed to encompass both projective measurement-based assertions and the Bloch vector-based assertions introduced by our approach.
Utilizing the depicted density matrix in \cref{fig:diagram_algorithmic_structure} as a foundation, we provide the following definition.

\begin{definition}[Runtime Assertion]
A runtime assertion is defined as a sequence of three steps performed in the following order:
\begin{enumerate}
    \item \textbf{Syntax}: $\texttt{Assert}\left( \mathcal{A} \right)$
    \item \textbf{Quantum Sampling}: $\mathcal{D} = \Big[\Tr ( \rho_{q,k} M_0), \Tr ( \rho_{q,k} M_1), \dots, \Tr ( \rho_{q,k} M_{d-1}) \Big]$
    \item \textbf{Predicate}: $\mathcal{A}$ is true $\iff \| \mathcal{D} - \mathcal{E}\| < \delta_{q,k}$
\end{enumerate}
\label{def:assertion}
\end{definition}

In \cref{def:assertion}, (1) defines the syntactic statement of the assertion at segment $k$ and qubit $q$ of the quantum program, representing a predicate statement $\mathcal{A}$ that assesses whether the measured density matrix $\rho_{q,k}$ is equal to the expected density matrix $\sigma_{q,k}$, within an error $\delta_{q,k}$.
Following this, (2) defines the quantum sampling operation to obtain classic data to assess $\mathcal{A}$ by storing the resulting data in the $d$-component vector $\mathcal{D} =\big[\mathcal{D}_0, \mathcal{D}_1,\cdots \mathcal{D}_{d-1}\big]$, where each data component $D_j$ contain the results from the measurement of an approach-specific operator $M_j$.
Then, in (3) we define the predicate $\mathcal{A}$ which is true when the difference between the measured data and the theoretical values $\mathcal{E}=\big[ \mathcal{E}_0, \mathcal{E}_1, \cdots, \mathcal{E}_{d-1} \big]$ is less than the error $\delta_{q,k}$, as computed by the approach-specific norm $\|\cdot\|$.

Following \cref{def:assertion}, projective measurements operate as follows: (1) the assertion is initiated in segment $k$ for qubit $q$; (2) projection operators, such as $\Tr(\rho_{q,k} P)$, are measured to obtain the corresponding measurement results; and (3) the classical data obtained in step (2) is evaluated to determine whether the measured basis state aligns with the expected outcome.

While assertions based on projective measurements offer strong logical expressive power, their execution on real quantum systems involves three steps: uncomputing the state, performing a mid-circuit measurement, and then recomputing the state. 
This process introduces additional gates and mid-circuit measurement operations, increasing the overall circuit depth—particularly problematic for \gls{nisq} devices. 
For example, adding a single assertion to a 10-qubit \gls{qft} circuit roughly triples the depth, due to the added uncompute and recompute operations.
After hardware-aware transpilation on IBM's \texttt{ibm\_sheerbroke} backend, the depth increases from 13 to 39.
\footnote{We measured this using Qiskit with default transpilation settings and a \gls{qft} circuit with one projective measurement assertion inserted.}

Projective measurement-based assertions rely heavily on mid-circuit measurements, which are available on certain quantum systems, such as IBM's superconducting backends (e.g., \texttt{ibm\_sherbrooke})~\cite{ibm_quantum_2021}. 
Mid-circuit measurements are essential for various quantum computing algorithms, including Shor's algorithm~\citep{Shor_1997}, and play a critical role in quantum error correction~\citep{rudinger2022characterizing}. 
However, these measurements are susceptible to errors, such as cross-talk, even when mitigation strategies like dynamical decoupling are employed~\citep{gaebler2021suppression}.

To avoid significant increases in circuit depth and reliance on mid-circuit measurements, we utilize assertions that measure the single-qubit density matrix at a specific segment and qubit, comparing it with the expected density matrix.
Given that the Bloch vector, introduced in \cref{eq:bloch_vector}, is equivalent to the density matrix of the corresponding state as defined in \cref{eq:density_matrix_representation_single_qubit}, it is sufficient to measure the Bloch vector. 
This vector is experimentally obtained by measuring the Pauli expectation values.

\begin{definition}[Bloch Vector Runtime Assertion]
A \textit{Bloch Vector Runtime Assertion} (Bloq assertion) $\mathcal{B}_{q,k}$ is defined as a sequence of three steps performed in order:

\begin{enumerate}
    \item \textbf{Syntax}: $\texttt{Assert}\left( \mathcal{B}_{q,k} \right)$
    \item \textbf{Quantum Sampling}: $\mathcal{D} = \Big[\Tr ( \rho_{q,k} X), \Tr ( \rho_{q,k} Y), \Tr ( \rho_{q,k} Z) \Big] = \texttt{Bloch Vector}$
    \item \textbf{Predicate}: $\mathcal{B}_{q,k}$ is true $\iff \| \mathcal{D} - \mathcal{E}\|_{\infty} < \delta_{q,k}$
\end{enumerate}
\label{def:bloq_assertion}

In \cref{def:bloq_assertion}, we define (1) as the syntactic statement of the Bloq assertion at segment $k$ and qubit $q$ of the quantum program, representing the predicate $\mathcal{B}_{q,k}: \rho_{q,k} = \sigma_{q,k}$.
This predicate evaluates whether the measured density matrix $\rho_{q,k}$ matches the expected density matrix $\sigma_{q,k}$ within a specified error $\delta_{q,k}$. 
With definition (2), we define the quantum sampling to obtain the Bloch vector from the expectation values of the Pauli matrices $X$, $Y$, and $Z$. 
Finally, (3) defines the predicate of the Bloq assertion as true if and only if the maximum norm difference between the sampled data $\mathcal{D}$ and the expected Bloch vector $\mathcal{E} = \big[\Tr(\sigma_{q,k} X), \Tr(\sigma_{q,k} Y), \Tr(\sigma_{q,k} Z)\big]$ is within the error threshold $\delta_{q,k}$. 
\end{definition}

With the design of Bloq, we insert assertions at every qubit in every segment of the quantum circuit. 
As with Proq, this yields a runtime efficiency that scales with the complexity of the quantum algorithm itself. 
For instance, Bloq maintains polynomial runtime for quantum algorithms with polynomial complexity. 
In contrast, the approach of \citet{Stat}, which relies on statistical tests over the entire output distribution, may introduce scalability challenges for algorithms with large output spaces, such as \gls{qft}, where Bloq remains efficient.

\subsection{Test Assessment}
\label{sec:approach_test_assessment}

For Bloq's test assessment, we utilize the quantum state fidelity, as defined in \cref{eq:state_fidelity}, between the sampled density matrix and the theoretical density matrix.
The sampled density matrix, denoted as $\sigma$, is reconstructed by sampling the expectation values of the $X$, $Y$, and $Z$ operators and substituting them into \cref{eq:density_matrix_representation_single_qubit}.
The theoretical density matrix, denoted as $\rho$, is derived from the algorithm-specific expected values produced by AutoBloq, as described in \cref{eq:density_matrix_representation_correct_state,eq:density_matrix_expected_grover}.  
Once both $\rho$ and $\sigma$ are obtained, the quantum state fidelity between them is computed using \cref{eq:state_fidelity}.

After computing the fidelity, we evaluate whether it exceeds a predefined test assessment threshold.  
This threshold is necessary for two reasons.  
First, it accounts for inherent statistical uncertainty resulting from finite sampling.  
Second, backend noise introduces systematic deviations in the measured distributions, requiring an additional tolerance to account for such noise-induced shifts~\citep{ShorErrorCorr,IBMAtScale}. 
If the computed fidelity exceeds the threshold, no fault is detected; otherwise, a fault is indicated at the corresponding circuit segment.

It is important to note that backend noise varies with circuit depth due to the accumulation of errors, primarily originating from two-qubit gates.  
For example, a noise-induced fidelity deviation of \SI{3}{\percent} may be typical, meaning that any assertion exhibiting a fidelity deviation greater than \SI{3}{\percent} from the expected value can be considered indicative of a fault.  
However, if the actual noise-induced deviation is \SI{4}{\percent} instead of the expected \SI{3}{\percent}, applying a threshold of \SI{3}{\percent} may result in the fault going undetected.

\begin{definition}[Bloq Single-Qubit Test Assessment]
Given a theoretical density matrix $\rho$ and a sampled density matrix $\sigma$, both of shape $2\times2$, corresponding to qubit $q$ at segment $k$.
Let $t$ be the test assessment threshold, where $t$ is a percentage between 0 and 100 (inclusive), and let $F$ denote the quantum state fidelity defined in \cref{eq:state_fidelity}.  
We then define the Bloq single-qubit test oracle as:

\begin{equation}
    \label{def:oracle_bloq}
f_{Bloq}(q, k) = \begin{cases}
    FAIL &  \text{if} \;\; F(\rho_{q,k}, \sigma_{q,k}) < 1 - t/100 \\
    PASS & \text{otherwise}
\end{cases}
\end{equation}

In \cref{def:oracle_bloq}, if the state fidelity is below the test assessment threshold $1-t/\SI{100}{\percent}$, the oracle returns \textit{FAIL}, indicating a potential fault at qubit $q$ in segment $k$.  
Otherwise, the assertion passes and the oracle returns \textit{PASS}.
\end{definition}

Once the Bloq test assessment for a single qubit has been defined, we extend this concept to combine single-qubit assertions into a segment-level assertion.  
This provides a higher level of abstraction for fault detection at the granularity of individual circuit segments.

\begin{definition}[Bloq Segment Test Assessment]

Let $f_{Bloq}(q, k)$ be the Bloq single-qubit test assessment, as defined in \cref{def:oracle_bloq}, for qubit $q$ at an arbitrary segment $k$ in a quantum program with $n$ qubits and $\mathcal{K}$ segments.
We then define the Bloq segment test assessment as follows:
\begin{equation}
    \label{def:oracle_bloq_segment}
f_{Bloq}(k) = \begin{cases}
    PASS &  \text{if} \;\; f_{Bloq}(q, k) = PASS \;\; \forall \;\; q \in \{1, 2, \dots, n\} \\
    FAIL &  \text{otherwise}
\end{cases}
\end{equation}
In \cref{def:oracle_bloq_segment}, $f_{Bloq}(k)$ represents the segment-level test assessment at segment $k$.  
For a given segment, the segment-level assertion passes only if all individual single-qubit test assessments $f_{Bloq}(q, k)$ pass for every qubit $q = 0, 1, \dots, n-1$.  
If any single-qubit test assessment fails, the segment-level test assessment returns \textit{FAIL}.
\end{definition}

\subsection{Automatic Assertion Generation with \texttt{AutoBloq}}
\label{sec:autobloq}

A common challenge in quantum software testing is the lack of readily available assertions, which often depend on program-specific knowledge and algebraic calculations to design assertion schemes~\citep{Proq,Stat,QECA2020,BuggySegments,honarvar2020property}. 
This issue becomes particularly significant when testing multiple inputs across numerous test cases. 
In such scenarios, assertions must be manually calculated for each input, either by hand or using computer algebra systems like Mathematica, and the resulting schemes must be stored for future use.

Instead, we propose \texttt{AutoBloq}, a method that automates the generation of the expected Bloch vector at a test point \((q,k)\) for a given input \(\ket{j} = \ket{j_0j_1\cdots j_{n-1}}\). 
We denote this expected Bloch vector as \(\Vec{R}^{\sigma}_{q,k}(\ket{j})\), where the superscript \(\sigma\) indicates that it corresponds to the algorithm's expected Bloch vector.

To construct \texttt{AutoBloq} for a specific quantum algorithm, a mathematical procedure is required. 
This involves using algebraic methods to manipulate the expression of the density matrix into the form given in \cref{eq:bloch_vector}, from which the expectation values can be extracted for the specified segment, qubit, and input parameters.
We outline a streamlined process for generating automated Bloq assertions (\texttt{AutoBloq}) for any input of the quantum program under test and at any segment, in accordance with the segment definition in \cref{sec:bloq_assertions_for_fault_loc}.

\begin{definition}[Automatic Bloq Assertion Scheme]
\label{def:autobloq}
Given a quantum algorithm $\texttt{ALG} = \texttt{ALG}_0 \texttt{ALG}_1 \cdots \texttt{ALG}_{\mathcal{K}-1}$, where each $\texttt{ALG}_j$ is a unitary operation representing an iterative step of the algorithm, and an initial state $\ket{\psi_{\text{init}}}$.
We obtain the automatic Bloq assertion scheme as follows:
\begin{enumerate}
    \item Compute the state after $k$ segments: 
    \[
    \sigma_k = \texttt{ALG}_k \ket{\psi_{\text{init}}}\bra{\psi_{\text{init}}} \texttt{ALG}_k^\dagger
    \]
    \item Obtain the reduced density matrix for qubit $q$:
    \[
    \sigma_{q,k} = \Tr_{\{x \mid x \neq q\}}(\sigma_k)
    \]
    \item Express $\sigma_{q,k}$ in Bloch vector form:
    \[
    \sigma_{q,k} = \frac{1}{2} \left( \mathbbm{1} + \braket{X}_{q,k} X + \braket{Y}_{q,k} Y + \braket{Z}_{q,k} Z \right)
    \]
\end{enumerate}
The resulting assertion scheme is defined as the Bloch vector $\Vec{R}^{(\sigma)}_{q,k} = \big[ \braket{X}_{q,k}, \braket{Y}_{q,k}, \braket{Z}_{q,k} \big]$.
\end{definition}

We now describe the steps required to compute the automated Bloq assertion scheme (\texttt{AutoBloq}).

\paragraph{Step 1: Compute the density matrix after $k$ steps.}
The first step calculates the density matrix \(\sigma_k\), representing the quantum state after \(k\) segments. 
This is done by evolving the initial state \(\ket{\psi_{\text{init}}}\) through the quantum gates \(\text{ALG}_0, \text{ALG}_1, \dots, \text{ALG}_k\) and their adjoints:
\[
\sigma_k = \text{ALG}_0 \text{ALG}_1 \cdots \text{ALG}_k \ket{\psi_{\text{init}}}\bra{\psi_{\text{init}}} \text{ALG}_k^{\dagger} \cdots \text{ALG}_1^{\dagger} \text{ALG}_0^{\dagger}.
\]
This encapsulates all transformations up to segment \(k\), preparing the state for subsequent operations.

\paragraph{Step 2: Compute the reduced density matrix for qubit \(q\).}
The reduced density matrix \(\sigma_{q,k}\) isolates the state information for qubit \(q\) by tracing out all other qubits:
\[
\sigma_{q,k} = \Tr_{\{ x \mid x \neq q \}}( \sigma_k ).
\]
This operation retains only the information relevant to qubit \(q\) at segment \(k\), forming the basis for constructing the Bloq assertion scheme.

\paragraph{Step 3: Express \(\sigma_{q,k}\) in Bloch vector form.}
The reduced density matrix \(\sigma_{q,k}\) is expressed in terms of its Bloch vector representation by decomposing it into a linear combination of the Pauli matrices \(X\), \(Y\), and \(Z\):
\[
\sigma_{q,k} = \frac{1}{2} \Big( \braket{X}_{q,k}(\ket{j}) X + \braket{Y}_{q,k}(\ket{j}) Y + \braket{Z}_{q,k}(\ket{j}) Z \Big).
\]
This representation allows direct extraction of the expectation values \(\braket{X}_{q,k}(\ket{j})\), \(\braket{Y}_{q,k}(\ket{j})\), and \(\braket{Z}_{q,k}(\ket{j})\), which define the Bloch vector.

\paragraph{Return: The Bloq assertion scheme.}
The Bloq assertion scheme is returned as the Bloch vector:
\[
\Vec{R}^{(\sigma)}_{q,k}\big(\ket{j}\big) = \big[ \braket{X}_{q,k}(\ket{j}), \braket{Y}_{q,k}(\ket{j}), \braket{Z}_{q,k}(\ket{j}) \big].
\]
This vector provides the expected Pauli operator values for qubit \(q\) at segment \(k\), given the input \(\ket{j}\), enabling automated testing at the specified segment.

By following these steps, \texttt{AutoBloq} provides a systematic and automated approach to generating assertions for quantum algorithms, significantly reducing the manual effort required for deriving assertion schemes.

\section{Evaluation}
\label{sec:evaluation}

In this section, we perform an experimental evaluation of Bloq.
We compare our assertions against projection-based assertions derived from the Proq approach~\citep{Proq}, which serves as our baseline in experiments conducted over two realistic quantum algorithms using both an ideal and a noisy quantum simulator. 
In total, we evaluate \num{684432} quantum program instances, spanning combinations of inputs, fault locations, fault types, and other experimental parameters. 
Details of both the experimental setup and baseline implementation are provided in the following sections.

\subsection{Research Questions}

Here, we present our research questions. Since noise is a critical factor in fault localization, we analyze all research questions under both ideal and noisy simulator conditions.

\begin{enumerate}
    \item \textbf{RQ1:}  
    How effective is Bloq's fault localization?  

    \item \textbf{RQ2:}  
    How does the number of qubits influence the effectiveness of Bloq's fault localization?

    \item \textbf{RQ3:}  
    To what extent does circuit depth impact Bloq's fault localization effectiveness?
    
    \item \textbf{RQ4:}  
    What is the impact of synthetic faults on Bloq’s effectiveness?
    
    \item \textbf{RQ5:}  
    What are the runtime and overhead costs of Bloq's fault localization? 
\end{enumerate}

As quantum program types have different structures in terms of gate composition and circuit depth, RQ1 establishes whether Bloq’s fault localization effectiveness varies for our two program types \gls{qft} and Grover under ideal, noisy conditions in comparison to the baseline.
RQ2 investigates how Bloq's performance scales with the number of qubits, a key determinant of circuit size and complexity.
Furthermore, since higher-depth circuits tend to amplify noise and error propagation, RQ3 investigates how circuit depth influences Bloq’s localization effectiveness. 
This RQ helps determine whether Bloq's approach is effective across varying depths.
In RQ4, we examine how the fault categories, add, remove, and replace, affect Bloq’s localization effectiveness. 
This helps us identify whether Bloq performs differently across categories and highlights strengths or limitations in detecting faults within each.

RQ5 explores the trade-offs between localization effectiveness and computational efficiency with respect to noise and in comparison to the baseline. 
Practical deployment of fault localization tools requires balancing accuracy with runtime and resource consumption.

\subsection{Study Subjects}
\label{sec:study_subjects}

In line with prior evaluations of runtime assertion methods~\citep{Proq,QECA2020}, which typically assess only a small number of representative quantum algorithms, we aim to select two widely studied and practically relevant algorithms: \gls{qft} and Grover’s search algorithm~\cite{QCQIBook,GroversAlgo}.  
For example, \citet{QECA2020} evaluate on \gls{qft} and quantum phase estimation, and additionally include Bernstein–-Vazirani.  
We exclude algorithms such as the latter due to their limited relevance for practical applications.

The \gls{qft} is exponentially faster than classical Fourier transforms~\citep{QCQIBook} and commonly used as a subroutine in key quantum algorithms such as Shor’s algorithm~\citep{Shor_1997} and the \gls{hhl} algorithm~\citep{HHL}.  
One of the primary reasons for selecting \gls{qft} is its ability to generate an exponentially large superposition when initialized with a computational basis state.
This ensures that our evaluation is not biased toward fault localization methods that rely on frequent sampling of the quantum register during execution~\citep{BuggySegments,AutomaticRepairQuantum}.  
By applying Bloq assertions to \gls{qft}, we assess Bloq’s ability to scale with qubit count and handle complex, highly entangled states.

Grover’s algorithm, on the other hand, is a direct application algorithm for solving the unstructured search problem.  
It produces a single output value with high probability, contrasting with the exponential superposition output of \gls{qft}.  
By including both algorithms, we evaluate Bloq across distinct classes of quantum programs—those with large output spaces and those with sharply peaked outputs—thereby demonstrating the generalizability and scalability of our approach.

While we focus on \gls{qft} and Grover in this evaluation, our approach is not limited to these two algorithms.  
It only requires that the target algorithm follow the structural assumptions described in \cref{sec:background_quantum_algorithms}, which apply to most quantum algorithms to the best of our knowledge.  
Thus, \gls{qft} allows us to assess Bloq's applicability to quantum programs that result in highly entangled, large superposition states.  

Since our evaluation is conducted on a noisy simulator using the noise model of IBM's \texttt{ibm\_sherbrooke} backend, the least noisy available at the time, we select qubit counts for \gls{qft} from \( q=2 \) to \( q=10 \) in incremental steps.
For circuits beyond 10 qubits, the backend ibm\_sherbrooke, accumulated noise from imperfect gate operations in the \texttt{ibm\_sherbrooke} backend introduces significant uncertainty, making fault localization results unreliable.
We evaluate \gls{qft} on all \(2^q\) computational basis inputs for each qubit count to ensure exhaustive coverage of input states.
For instance, at \( q=2 \), we assess all four possible inputs $\ket{00}$, $\ket{01}$, $\ket{10}$ and $\ket{11}$, and at \( q=3 \), we evaluate all eight possible inputs $\ket{000}$, $\ket{001}$, ..., $\ket{111}$.  
This approach ensures that Bloq's performance is tested across the full range of input states constrained by the qubit count.  

The second algorithm we include is Grover’s algorithm, which yields a single, well-defined output state, in contrast to \gls{qft}, which produces a superposition over \(2^q\) basis states. 
This distinction matters because many testing approaches require a classical oracle to evaluate all outputs~\cite{Stat,BuggySegments,AutomaticRepairQuantum}, which becomes infeasible for large superpositions. 
Our method avoids this by using Bloq assertions on local qubit properties such as the Bloch vector, making it suitable for both small and large output distributions at scale.
By evaluating both \gls{qft} and Grover, we capture the extremes of the distributions and implicitly cover intermediate cases as well.
 
Furthermore, unlike \gls{qft}, Grover’s algorithm has a significantly greater circuit depth, which makes it more susceptible to noise accumulation.  
Since circuit depth directly influences error propagation, we expect Grover’s performance to degrade faster as qubit count increases.  
For this reason, we evaluate Grover’s algorithm at qubit counts ranging from \( q=2 \) to \( q=6 \) incrementally.  
Beyond \( q=6 \), from findings in our trial runs, the depth-induced noise accumulation is expected to be too high to yield meaningful results, making further evaluation impractical under the noise constraints.
Due to Grover's high scaling in circuit depth, this is a common challenge in the research field~\cite{ProspectGrover_Wang}.

As with \gls{qft}, we evaluate Grover's algorithm across all possible computational basis inputs constrained by the qubit count, ensuring a complete assessment of Bloq’s fault localization across different input states.  

By selecting these two algorithms and setting qubit count limits based on noise constraints, we aim to capture a representative range of quantum program behaviors, from broad superpositions to narrow outputs—under realistic noisy conditions.
While we do not evaluate Bloq on all types of quantum programs, \gls{qft} and Grover span the extremes of output structure and distribution, offering meaningful diversity for assessing effectiveness. 
We exclude algorithms that do not rely on quantum gate-based circuits, such as quantum annealing, which instead operate through adiabatic evolution of a physical system to reach a low-energy solution~\cite{FalcoAnnealing}.  
We also exclude hybrid variational algorithms like \gls{qaoa} and \gls{vqe}, which rely on measuring intermediate expectation values which are passed on to the next iterative step in the algorithm to update quantum gate parameters.  
Bloq’s current framework does not support passing measured values between iterations, although we discuss how such support could be added in future work.

\subsection{Experiment Design and Setup}
\label{sec:experiment_design_and_setup}

\begin{figure*}[tbp]
    \centering
    \includegraphics[width=\textwidth]{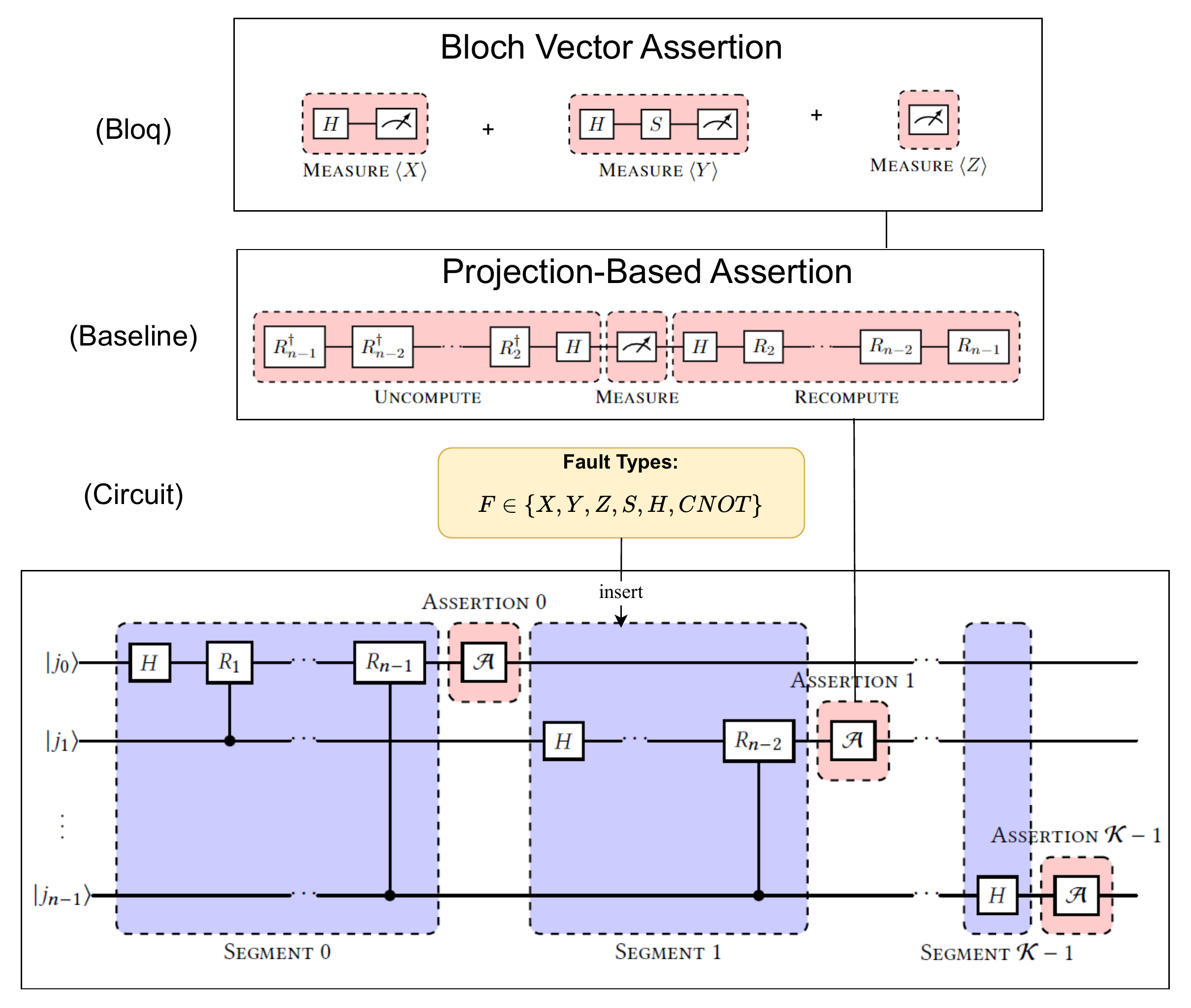}
    \caption{Overview of our evaluation design for \gls{qft}. The diagram shows how Bloq assertions (top) and projection-based assertions (Proq, middle) are inserted into circuit segments (bottom). Faults $F \in \{X, Y, Z, S, H, \text{CNOT}\}$ are injected into the segments. Bloq assertions measure Bloch vector components $\{\braket{X}$, $\braket{Y}$, $\braket{Z}\}$), while Proq assertions follow the \textbf{UNCOMPUTE}–\textbf{MEASURE}–\textbf{RECOMPUTE} process described in \cref{sec:background_projection_schemes}.}
    \label{fig:experiment_design_diagram}
\end{figure*}

\Cref{fig:experiment_design_diagram} illustrates the structure of our experiment design.  
At the bottom of the figure, we depict a quantum circuit labeled \textit{Circuit}, which represents a general \gls{qft} program with $n$ qubits and $\kappa$ segments.  
At each segment, we insert assertions, denoted by $\mathcal{A}$. These assertions can be either of type (Bloq) or (Proq), as shown in the upper part of \cref{fig:experiment_design_diagram}.
For \gls{qft}, there is only one possible position for an assertion within each segment. 
Therefore, we insert assertions at the specific qubit according to the order in which they are operated on.  
In contrast, for the Grover program type, we insert one assertion per qubit in each segment, resulting in the same number of assertions per segment as there are qubits in the program.

In both cases, we insert either Bloq or Proq assertions, following their respective assertion schemes, and compare their effectiveness at detecting faulty segments.

\subsubsection{Assertion Schemes}
\label{sec:eval_design}

We use projection-based assertions (Proq) as our baseline, as illustrated in the second panel, from the top, of \cref{fig:experiment_design_diagram}.  
As was introduced in \cref{sec:background_projection_schemes}, Proq follows a three-stage process: \textbf{UNCOMPUTE}, \textbf{MEASURE}, and \textbf{RECOMPUTE}.  
For \gls{qft}, we implement the \textbf{UNCOMPUTE} operation using the inverse of the single-qubit gates defined in \cref{eq:qft_transformation}, applied to the specific qubit operated on in each segment.  
This simplification avoids two-qubit gates by leveraging knowledge of the input state.  
For Grover, we apply the full inverse of the Grover operator in the \textbf{UNCOMPUTE} stage and reapply it for \textbf{RECOMPUTE}, as its entangled structure prevents qubit-level isolation.  
Measurement results from each assertion are stored in separate classical registers and compared to the expected input bits to assess Proq’s fault detection capability.

The top panel of \cref{fig:experiment_design_diagram} shows our Bloq assertion scheme, which evaluates the Bloch vector components of each qubit by measuring along the \(X\), \(Y\), and \(Z\) axes.  
Both assertion types are inserted into each segment of the program under test, as illustrated in the lower portion of \cref{fig:experiment_design_diagram}.  
Assertions are placed after the relevant gates in each segment, and their outcomes are used to detect the presence and location of injected faults.

For the Bloq assertions, no additional engineering of the assertions is required. 
Instead, we utilize AutoBloq, as described in \cref{def:autobloq}, to automatically compute the assertion schemes for both \gls{qft} and Grover programs, as we perform in the coming sections.
As shown in (Bloq) in \cref{fig:experiment_design_diagram}, Bloq assertions insert three individual gates to compute the expectation values $\braket{X}$, $\braket{Y}$, and $\braket{Z}$. 
These expectation values, in turn, form the Bloch vector at the corresponding circuit location, which we collect during test.
Thus, for any program type, Bloq assertions always follow the same form, in contrast to projection-based assertions, which require specialized insertion gates depending on the quantum algorithm and the specific engineering of the assertions.
In the following sections, we apply AutoBloq to generate the Bloq assertion schemes used in our experiments for both \gls{qft} and Grover.

We utilize Qiskit's \texttt{Estimator} primitive to perform Bloq assertions by estimating the expectation values $\braket{X}$, $\braket{Y}$, and $\braket{Z}$.  
Under the hood, the Estimator automatically inserts basis change gates before measurement:  
$\braket{X}$ is obtained by applying a Hadamard gate prior to measurement;  
$\braket{Y}$ requires a Hadamard followed by an $S$ gate;  
$\braket{Z}$ is measured directly in the computational basis without additional gates.

\subsubsection{Quantum Fourier Transform}
\label{sec:autobloq_qft}

We now apply the AutoBloq method, introduced in \cref{def:autobloq}, step by step to the \gls{qft}.

Starting from Step 1, we consider the product form of the \gls{qft} transformation as given in \cref{eq:qft_transformation}. 
From this, the state of a single qubit can be expressed as:

\begin{equation}
   \ket{\theta}= \frac{1}{\sqrt{2}}\Big( \ket{0} + e^{i\theta} \ket{1}\Big) = R_z(\theta) H \ket{0}
   \label{eq:theta_state}
\end{equation}

In \cref{eq:theta_state}, the transformation $ALG_k = R_z(\theta) H$, as the iterative structure of the \gls{qft} algorithm applies operations qubit by qubit in isolation—fully completing all operations on the first qubit before proceeding to the next. 
Accordingly, for \gls{qft}, the segment index $k$ corresponds directly to the qubit being operated on.
The rotation operator in \cref{eq:theta_state} is defined as:

\begin{equation}
    R_z(\theta) = \begin{pmatrix}
        1 & 0\\
        0 & e^{i\theta}
    \end{pmatrix}
\end{equation}

To complete Step 2 of \cref{def:autobloq}, we apply the density matrix representation from \cref{eq:density_matrix_representation_single_qubit} and express the non-faulty state from \cref{eq:theta_state} as:

\begin{equation}
    \rho = \ket{\theta}\bra{\theta}= \frac{1}{2}\Big( \mathbbm{1} + \cos\theta\, X + \sin\theta\, Y \Big)
\label{eq:density_matrix_representation_correct_state}
\end{equation}

In \cref{eq:density_matrix_representation_correct_state}, we use the identities $\braket{X} = \cos\theta$, $\braket{Y} = \sin\theta$, and $\braket{Z} = 0$. 
To complete Step 3, we extract the expected values directly from \cref{eq:density_matrix_representation_correct_state}, corresponding to the coefficients of the $X$, $Y$, and $Z$ operators, respectively.

This yields the following Bloq assertion scheme for \gls{qft}:

\begin{equation}
    \texttt{AssertionScheme}_{q,k}(j) = \Big[ \cos\theta_{q,k}(j), \sin\theta_{q,k}(j), 0 \Big]
    \label{eq:bloq_qft_assertion}
\end{equation}

In \cref{eq:bloq_qft_assertion}, we substitute $\theta$ with $\theta_{q,k}(j)$ from \cref{eq:qft_transformation}, which depends on the qubit $q$, the segment index $k$, and the input $j$. 
For example, in the case of $n = 2$ qubits (i.e., $k = 2$), and an input $j = 01$, we obtain $\theta_{2,2}(01) = 2\pi \cdot 0.10$.

In our experiments, whenever we compute a Bloq assertion, we refer to \cref{def:bloq_assertion}. 
Specifically, the result of \cref{eq:bloq_qft_assertion} defines the theoretical Bloch vector $\sigma_{q,k}$ in that definition, corresponding to qubit $q$ and segment $k$. 
This allows us to generate Bloq assertion schemes on demand for any program input $j$, which we then compare to the measured density matrix $\rho_{q,k}$ utilizing our test assessment defined in \ref{sec:approach_test_assessment}.

    \subsubsection{Grover's Algorithm}
    \label{sec:autobloq_grover}
    
    To perform Step 1 of AutoBloq in \cref{def:autobloq}, we define the segment operator $ALG_k$ as the Grover operator from \cref{eq:grover_operator}. 
    Accordingly, the state after $k$ Grover iterations is given by:
    
    \begin{equation}
        G^k \ket{\psi} = \cos\phi_k \ket{\alpha} + \sin\phi_k \ket{\beta}
        \label{eq:grover_iteration_k}
    \end{equation}
    
    Here, $\phi_k = \frac{2k+1}{2}\theta$, and the states $\ket{\alpha}$ and $\ket{\beta}$ represent the equal superpositions of non-solution and solution states, respectively, defined as~\citep{QCQIBook,GroversAlgo}:
    
    \begin{equation}
        \ket{\alpha} = \frac{1}{\sqrt{N - M}} \sum_j \ket{j}
        \label{eq:grover_non_solutions}
    \end{equation}
    
    \begin{equation}
        \ket{\beta} = \frac{1}{\sqrt{M}} \sum_j \ket{j}
        \label{eq:grover_solutions}
    \end{equation}
    
    We restrict the analysis to the case of a single marked item, $M = 1$, which corresponds to the baseline version of Grover's algorithm. 
    This case is commonly used in both theoretical analyses and practical demonstrations, and it is representative of the typical unstructured search scenario.
    With this choice of $M$, we simplify \cref{eq:grover_solutions} down to:
    
    \begin{equation}
        \ket{\beta} = \ket{j_1j_2\cdots j_n}
    \end{equation}
    
    For Step 2 of \cref{def:autobloq}, we compute the full density matrix after $k$ Grover iterations as:
    
    \begin{align}
        \rho_k &= \cos^2\phi_k \ket{\alpha}\bra{\alpha} + \cos\phi_k \sin\phi_k \ket{\alpha}\bra{\beta} \\
        &+ \cos\phi_k\sin\phi_k\ket{\beta}\bra{\alpha} + \sin^2\phi_k \ket{\beta}\bra{\beta}
    \end{align}
    
    In Step 3, we perform a partial trace over all qubits except a selected qubit $q$. 
    The detailed derivation of this partial trace is provided in \cref{appendix:grover_calculation}; here, we state the resulting reduced density matrix:
    
    \begin{align}
        \rho_{k, q} &= \frac{1}{2}\Big( \mathbbm{1} + \Big[ \frac{\cos^2\phi_k}{N-1}(N-2)+\frac{\sin 2\phi_k}{\sqrt{N-1}} \Big]X\\
        &+\Big[ \sin^2\phi_k - \frac{\cos^2\phi_k}{N-1} \Big](-1)^{j_q}Z\Big)
        \label{eq:density_matrix_expected_grover}
    \end{align}
    
    From \cref{eq:density_matrix_expected_grover}, we observe that the expectation value of the $Y$ operator is zero, while the expectation values of $X$ and $Z$ depend on the segment index, the qubit $q$, and the input $j$.
    
    As in the case of \gls{qft}, we extract the expected values directly from \cref{eq:density_matrix_expected_grover}, resulting in the following assertion scheme for Grover:
    
    \begin{equation}
        \texttt{AssertionScheme}_{q,k}(j) = \Big[ \braket{X}_{q,k}(j), 0, \braket{Z}_{q,k}(j) \Big]
        \label{eq:grover_assertion_scheme}
    \end{equation}
    
    In \cref{eq:grover_assertion_scheme}, the expectation values are given by:
    
    \begin{equation}
        \braket{X}_{k,q} =  \frac{\cos^2\phi_k}{N-1}(N-2)+\frac{\sin 2\phi_k}{\sqrt{N-1}}
    \end{equation}
    
    \begin{equation}
        \braket{Z}_{k, q} = \Big[ \sin^2\phi_k - \frac{\cos^2\phi_k}{N-1} \Big](-1)^{j_q}
    \end{equation}
    
    As shown, the expectation values depend on the qubit index $q$, the segment index $k$, and the input $j$.
    The input dependency is encoded through $j_q$, which denotes the value of input $j$ at qubit position $q$. 
    For example, if the input is $j = 01$ and $q = 1$, then $j_q = j_1 = 1$.

In our experiments, the result of \cref{eq:grover_assertion_scheme} corresponds to the theoretical density matrices $\rho_{q,k}$ in \cref{def:bloq_assertion}, for qubit $q$ and segment $k$. 
This allows us to generate Bloq assertion schemes on demand for any program input $j$, which we then compare to the measured density matrix defined in \ref{sec:approach_test_assessment}.

\subsubsection{Experiment Setup}

In this section, we detail the experimental parameters of our evaluation and explain how we control for them.

\paragraph{\textbf{Segment Count}}

Each quantum program consists of a fixed number of segments, which depends on the number of qubits in the program.  
For \gls{qft}, the number of segments is equal to the qubit count, since the iterative step in the \gls{qft} algorithm operates on each qubit individually, except for the controlled operations.  
Thus, for \gls{qft}, the segment count ranges from \num{2} to \num{10}.
For Grover, the number of segments also depends on the number of qubits.  
However, unlike \gls{qft}, Grover's iterative step is determined by the number of repetitions of the Grover operator, as defined in \cref{eq:grover_operator}.  
Grover's algorithm provides a theoretical formula for calculating the optimal number of repetitions of the Grover operator, given by $\left\lfloor \frac{\pi}{4} \sqrt{2^n} \right\rfloor$~\citep{GroversAlgo}.
Using this formula, we determine the number of segments to be \{1, 2, 3, 4, 6\} when evaluated over the qubit range selected in \cref{sec:study_subjects}.

\paragraph{\textbf{Assertion Count}}

The general rule we follow is to perform assertions immediately after each segment, at the corresponding circuit location.
This follows directly from our definition of a segment as the code executed between two iterative steps in the quantum algorithm.  
Given this definition, placing assertions before a segment would contradict causality, as the assertion would occur prior to the code it is meant to evaluate.  
Thus, placing assertions after each segment is not just a design choice but a structural necessity.  
Alternative definitions of segments, for instance, based on fixed gate counts or circuit depth, could permit different assertion placements, but would depart from the algorithmic structure we aim to exploit for testing and debugging.
Following a segment, we insert either Bloq assertions or the baseline projection-based (Proq) assertions, depending on the evaluation scheme.

For Bloq assertions, we perform measurements of the $X$, $Y$, and $Z$ expectation values on each individual qubit in the circuit. 
This approach is applied to both \gls{qft} and Grover programs.
Thus, the number of assertions required for any given circuit is equal to the number of qubits connected to a segment, multiplied by the total number of segments.  
For our evaluation of \gls{qft}, this results in an assertion count that ranges from 2 to 10, corresponding directly to the number of qubits.  
For Grover, the assertion counts are \{2, 6, 12, 20, 36\}, as determined by the combination of segment counts and qubit counts used in our experiments.

For the baseline approach, Proq, we apply single-qubit assertions for the \gls{qft} programs. 
This results in the same assertion count as for the Bloq approach, since assertions are performed on each individual qubit following every segment.  
In contrast, for Grover, we use multi-qubit assertions in the baseline approach. 
Here, a single assertion is performed per segment, resulting in an assertion count that corresponds directly to the number of segments. 
Consequently, the range of assertion counts for Grover under the baseline approach matches the range of segment counts defined earlier.

\paragraph{\textbf{Circuit Shots}}

For each sampling execution of a quantum program, we perform \num{8192} shots.  
We selected \num{8192} shots because it is the maximum number allowed by any Qiskit backend at the time of writing of this paper, enabling us to achieve the highest possible precision in our probability estimates~\citep{QiskitCite}.

With this level of precision selected, we now account for the number of shots required by the Bloq approach compared to the baseline approach.

For the baseline approach, the assertions are integrated into the circuit and executed as part of the circuit run. Therefore, we require only \num{8192} shots for each execution of the baseline approach.

In contrast, for Bloq, we require $3 \times \num{8192}$ shots per single-qubit Bloq assertion, as each assertion involves measuring the expectation values $\braket{X}$, $\braket{Y}$, and $\braket{Z}$ separately.
We terminate the execution of Bloq once a fault has been detected. 
This implies that we do not necessarily need to execute all Bloq assertions, but only as many as are required to reach a test result, making the test localization more efficient.

\paragraph{\textbf{Fault Types}}

As of this writing, there have been two empirical studies of real quantum bugs~\citep{BugsinQCEmpirical,campos2021qbugs}.  
While these benchmarks are neither widely accepted nor easy to use due to a lack of interface against the faults and lack of fully faulty programs, as pointed out by \citet{fortunatoverification}, they represent the current state-of-the-art in quantum software testing.
Although these shortcomings may be restrictive in certain cases where an interface towards bugs or fully faulty programs is required, we argue that these benchmarks do show evidence that quantum-specific faults are present in relevant quantum programs today.
These benchmarks have already been applied to \gls{qst} techniques and are widely cited in the literature~\cite{repairGPT,AutomaticRepairQuantum,usandizaga2023quantum,ramalho2024testing,de2024quantum,zhao2023qchecker}.
In particular, \citet{BugsinQCEmpirical} found that \SI{39.9}{\percent} of 223 bugs were quantum-specific, with over a third producing incorrect outputs.  
Similarly,~\citep{Bugs4Q} reported that \SI{41.7}{\percent} of their manually validated Qiskit bugs led to incorrect outputs without crashing the program.
These output faults are particularly challenging to detect, as the program does not crash but silently produces incorrect results.  
They are often caused by incorrect implementations of quantum gates within the algorithm.

Based on this evidence, we evaluate our approach using faulty programs with gate faults from the following common gate set

\begin{equation}
    \{X, Y, Z, CNOT, S, H\}
\label{eq:gate_fault_set}
\end{equation}

, as these gates are abundantly applied in some of the most important quantum algorithms: QFT, Grover, Shor, QPE, and \gls{hhl}~\citep{QCQIBook,GroversAlgo,Shor_1997,HHL}.
Furthermore, we argue that these gates constitute a representative sample of
Quantum operations include common fundamental operations such as single-qubit bit flips, uniform superposition, phase flips, relative phase rotations, and two-qubit controlled operations, which quantum programmers frequently interact with~\citep{baltes2021sampling,abcSoftwareResearch}.

\paragraph{\textbf{Fault Category}}

In addition to the fault types described in the previous paragraph, we evaluate three fault categories in our experiments.  
These categories are:
\begin{inparaenum}
    \item \addgate,
    \item \removegate, and
    \item \replacegate.
\end{inparaenum}
We selected these categories because they are commonly studied in quantum software testing~\citep{Muskit,QMutPy}, and they represent the most common coding operations introduced either manually by programmers or automatically by compilers and optimization processes~\cite{QCQIBook,GroversAlgo,Venegas_Andraca_2012_Quantum_Walk,Proq,Stat,QECA2020}.
For the \addgate{} and \replacegate{} categories, we introduce faults by inserting gates from the fault types described in the previous paragraph. 
For the \removegate{} category, we delete an existing gate from the circuit at a segment location defined by our parameter set.

Specifically, for each fixed program segment, we perform an \addgate{} mutation for every fault type in our gate fault set (defined in \cref{eq:gate_fault_set}). 
For the \removegate{} category, we remove an arbitrarily selected gate within the segment. 
For the \replacegate{} category, we substitute an arbitrarily selected gate with a different one from the gate fault set. 
We ensure that the inserted gate differs from the one removed to properly simulate a fault.

\subsection{Test Assessment}
\label{sec:evaluation_test_assessment}

In our evaluation, we utilize the Bloq segment test assessments, as defined in \cref{def:oracle_bloq_segment}, to perform fault detection at each segment.  
Test assessments are conducted individually for each segment $k$, evaluating whether any faults are detected based on the combined results of the single-qubit assessments.
For the test assessment thresholds introduced in \cref{sec:approach_test_assessment}, we apply a range of thresholds from \SI{0}{\percent} to \SI{25}{\percent}, in \SI{1}{\percent} increments, to explore different tolerance levels for fault detection sensitivity.

For the baseline approach (Proq), we perform test assessments by comparing the sampled distributions obtained from projective measurements with the corresponding theoretical distributions.  
The comparison is conducted after performing mid-circuit projective measurements at each segment, and deviations beyond the specified threshold are used to determine the presence of a fault.
As with Bloq, a test assessment threshold $t$ is applied in the baseline (Proq) approach to account for sampling uncertainty and backend noise.

For \gls{qft}, the expected state distribution consists of a single computational basis state, which corresponds to the input state of the \gls{qft} circuit.  
For example, if the input to a 5-qubit \gls{qft} program is the state $\ket{01101}$, then the theoretical distribution expected after applying the assertions consists solely of the state $\ket{01101}$ with probability 1.

For Grover, however, since we apply Grover operators prior to performing the measurement, the expected theoretical state is $\ket{0}^{\otimes n}$, corresponding to the initial state after amplitude amplification and inversion have been applied.

\subsection{Metrics and Statistical Analyses}
\begin{table}[tbp]
    \centering
    \caption{Metrics and statistical methods utilized for each research question. 
    We apply Mann-Whitney U (MWU) tests, Confidence Interval (CI), Interquartile Range (IQR), and Vargha--Delaney effect size (\(\hat{A}_{12}\)) as shown. 
    The CI half-width refers to half the distance between the upper and lower CI bounds, representing the uncertainty around the mean.}
    \label{tab:eval_overview}
    \begin{tabular}{l l l}
        \toprule
        \textbf{Research Question} & \textbf{Metric} & \textbf{Statistic} \\
        \midrule
        RQ1 & F1 score & \(\hat{A}_{12}\), MWU, p-value, Mean, CI, CI half-width \\
        RQ2 & F1 score & \(\hat{A}_{12}\), MWU, p-value \\
        RQ3 & F1 score & \(\hat{A}_{12}\), MWU, p-value \\
        RQ4 & F1 score & \(\hat{A}_{12}\), MWU, p-value \\
        RQ5 & Runtime, Depth Overhead & Mean, Median, IQR \\
        \bottomrule
    \end{tabular}
\end{table}
Here, we detail the metrics and statistical analyses used in our evaluation, as summarized in \cref{tab:eval_overview}.

\subsubsection{Effectiveness Metric}

Given the nature of our test assessment results, each assertion outcome can be classified as either a positive or a negative.  
In this context, a positive outcome occurs when the Bloq segment test oracle, defined in \cref{def:oracle_bloq_segment}, returns \textit{FAIL}, indicating a potential fault at the corresponding segment.  
Conversely, a negative outcome occurs when the oracle returns \textit{PASS}, indicating no fault was detected.
  
These outcomes can be further categorized as true or false, depending on whether they correctly reflect the presence or absence of an injected fault:  
\begin{itemize}
    \item \textbf{True Positive (TP)}: A fault is present, and the oracle correctly returns \textit{FAIL}.
    \item \textbf{False Positive (FP)}: No fault is present, but the oracle incorrectly returns \textit{FAIL}.
    \item \textbf{True Negative (TN)}: No fault is present, and the oracle correctly returns \textit{PASS}.
    \item \textbf{False Negative (FN)}: A fault is present, but the oracle incorrectly returns \textit{PASS}.
\end{itemize}

Based on these classifications, we compute the F1 score to evaluate the fault detection performance.

In addition, since our goal is to evaluate the general effectiveness of our approach, we do not assign greater importance to either precision or recall.  
We therefore employ the F1 score to measure the predictive performance of our fault detection approaches~\citep{f1score,taha2015metrics}.  
The F1 score is defined as:

\begin{equation}
    F1(TP, FP, FN) = \frac{2TP}{2TP + FP + FN}
    \label{eq:f1_metric}
\end{equation}

In \cref{eq:f1_metric}, the F1 score is computed from the number of true positives ($TP$), false positives ($FP$), and false negatives ($FN$) of a particular grouping of the results.
To evaluate the particular scope of the research questions RQ1-RQ4: we compute the F1 score across program type for RQ1, inputs for RQ2, fault type and fault category for RQ3, and inputs and fault type for RQ4.
The F1 score provides a balanced measure of precision and recall, making it an appropriate choice for evaluating overall fault detection performance in our experiments.

\subsubsection{Test Runtime}

Test runtime, in our evaluation, is defined as the time required to reach a test result.  
For Bloq, since the approach operates segment by segment, the test runtime includes only the execution time necessary to identify a fault.  
In cases where a fault is detected early in the circuit, Bloq allows us to terminate the test without executing the remaining segments, thereby reducing the overall runtime.  
Specifically, the circuit is executed only up to the segment where the fault is detected, and no additional executions are required beyond that point.
Therefore, if a fault is inserted into an early segment, it can be detected quickly, requiring minimal circuit execution.  
Conversely, if a fault is inserted into the final segment, the test must execute the entire circuit up to and including that segment before the fault is detected.

For the baseline approach (Proq), however, the segments cannot be executed independently.  
The entire circuit, including the inserted \textbf{COMPUTE}, \textbf{MEASURE}, and \textbf{RECOMPUTE} operations described in \cref{sec:experiment_design_and_setup}, must be executed in full for each test assessment, regardless of where the fault occurs.

\subsubsection{Depth Overhead}

For both Bloq and Proq, a certain number of additional gates must be inserted into the program under test in order to perform test assessments.  
These inserted gates incur a depth overhead, increasing the overall circuit length and requiring additional runtime to complete execution.  
In addition, the increased circuit depth results in greater noise accumulation from quantum gates, which negatively impacts the overall circuit fidelity and measurement quality.
To quantify this overhead, we measure the physical circuit depth required to reach a test result for a given program under test.

\subsubsection{Statistical Analysis}

In addition to utilizing aggregate statistics such as the mean, \gls{iqr}, and bootstrapped \gls{ci}, we apply the \gls{mwu} test to assess the statistical significance of our results. 
We use bootstrapped confidence intervals to account for non-normal distributions, where standard deviations or standard errors would not provide reliable uncertainty estimates.
Furthermore, we compute the Vargha-Delaney effect size, denoted by $\hat{A}_{12}$~\citep{ArcuriBriand,VarghaDelanay}, to evaluate the strength of the observed effects identified by the \gls{mwu} test.
An effect size of $\hat{A}_{12} = 0.5$ indicates that there is no difference between the two approaches being compared.  
Values of $\hat{A}_{12} > 0.5$ indicate that the first approach outperforms the second, while values of $\hat{A}_{12} < 0.5$ favor the second approach.

To interpret the magnitude of the effect size, we use the scaled statistic $\hat{A}_{12}^{scaled} = 2(\hat{A}_{12} - 0.5)$, following established guidelines~\citep{HessRobustConfidence,laaber2024evaluating}. 
Based on the absolute value of this scaled effect size, we categorize the magnitude into four nominal groups:  
\begin{enumerate}
    \item \textit{Negligible} (N): $|\hat{A}_{12}^{scaled}| < 0.147$
    \item \textit{Small} (S): $0.147 \leq |\hat{A}_{12}^{scaled}| \leq 0.33$
    \item \textit{Medium} (M): $0.33 < |\hat{A}_{12}^{scaled}| < 0.474$
    \item \textit{Large} (L): $|\hat{A}_{12}^{scaled}| \geq 0.474$
\end{enumerate}

Results are considered statistically significant if the p-value is less than or equal to 0.05 and the effect size magnitude is greater than \textit{Negligible} (N).

\subsection{Threats to Validity}

In this section, we discuss potential threats to the validity of our evaluation, categorized as construct, internal, and external validity threats~\citep{Threats_to_Validity_Dag}.

\paragraph{Construct Validity}

Construct validity concerns whether the evaluation accurately measures the concepts it intends to assess. 
In our evaluation, the key construct is the effectiveness of assertion-based fault detection in quantum programs.  
One potential threat lies in the use of simulated faults, which may not fully represent the diversity or complexity of faults found in real-world quantum software. 
We mitigated this threat by selecting fault types based on existing empirical studies of quantum bugs~\citep{BugsinQCEmpirical, campos2021qbugs}, ensuring that our fault models reflect realistic scenarios.

Another potential threat concerns the engineering of the baseline assertion approach (Proq), which is based on projective measurements. 
In our evaluation, we implemented a specific assertion scheme for Proq, designed using Grover operators for the Grover algorithm. 
While this provides a reasonable and consistent comparison, it is possible that alternative assertion schemes could offer improved efficiency or effectiveness.  
To mitigate this threat, we applied a comparable engineering strategy to Bloq by designing its assertions using Grover operators as well. 
This alignment ensures that both approaches leverage equivalent algorithm-specific structures in their assertion schemes, thereby providing a fair basis for comparison.
  
Finally, we recognize that quantum state fidelity may not capture every aspect of assertion effectiveness. However, we selected fidelity as it provides a standardized and interpretable metric for quantifying the deviation between expected and observed quantum states~\cite{Jozsa01121994,QCQIBook,Proq}. 
Additionally, we performed a sensitivity analysis by varying the test assessment thresholds to explore different levels of detection strictness.

\paragraph{Internal Validity}

Internal validity refers to the extent to which the observed results are attributable to the experimental factors under our evaluation, rather than to confounding variables.  
In our case, backend noise and hardware variability may introduce confounding effects in fault detection outcomes. 
While we control for some sources of variability by using predefined test assessment thresholds and performing experiments over multiple runs, noise levels vary with circuit depth and qubit connectivity, which can affect fault detection rates.  
To mitigate these threats, we conducted experiments on both ideal (noise-free) and noisy quantum backends to distinguish between errors caused by injected faults and those caused by hardware noise. We also repeated experiments to reduce the impact of random fluctuations and ensure reproducibility of results.

\paragraph{External Validity}

External validity addresses the generalizability of our findings beyond the specific context of our evaluation.  
Our evaluation is limited to two quantum algorithms, \gls{qft} and Grover, and the experiments were conducted on a specific set of quantum hardware backends. 
While these algorithms are representative of distinct quantum computational patterns (Fourier transforms and amplitude amplification), they may not capture the full spectrum of behaviors in quantum programs.  
Additionally, our experiments are constrained by current hardware limitations, such as the number of available qubits and noise characteristics of \gls{nisq} devices. 
Despite the efficient runtime complexity of our approach, as quantum hardware evolves, the applicability of our results to larger systems and different architectures will need to be reevaluated.  
Nonetheless, we selected well-established algorithms and backends to provide a meaningful and representative reference for assessing assertion-based fault detection approaches in current quantum computing environments.

\section{Results}
\label{sec:results}

Here, we present the results from our evaluation.

\subsection{RQ1: Overall Effectiveness By Program Type}

\begin{figure}
    \centering
    \includegraphics[width=0.9\linewidth]{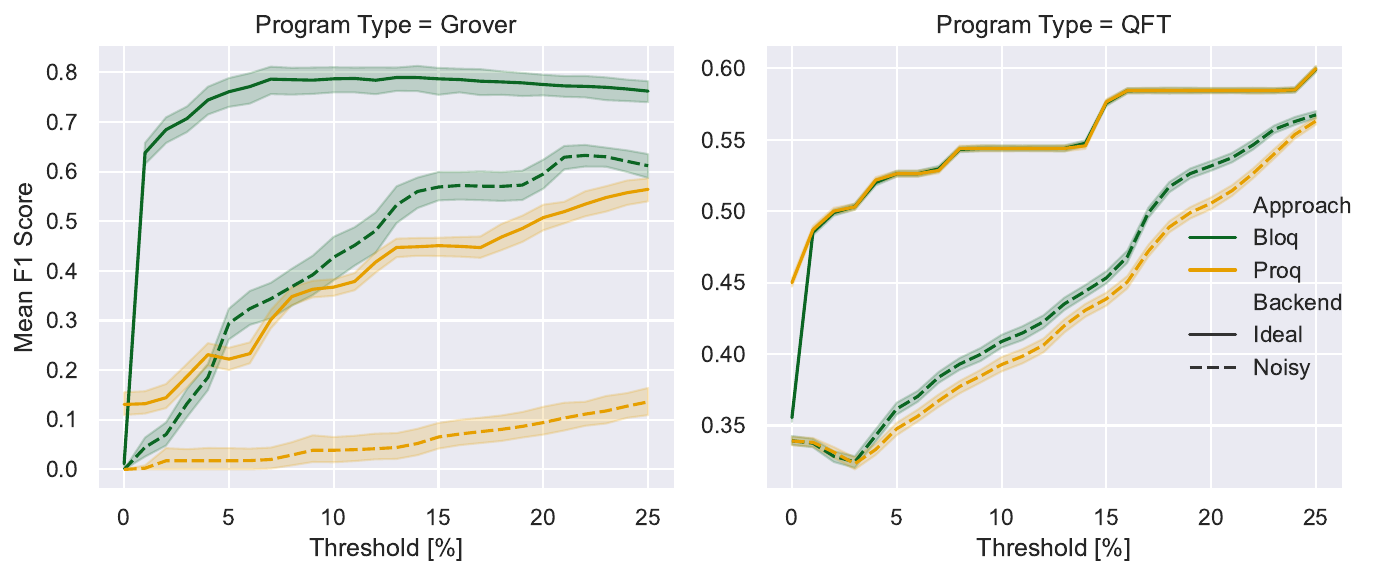}
    \caption{Mean F1 scores with bootstrapped \SI{99}{\percent} \gls{ci} as a function of the assertion deviation threshold (x-axis, in percent) for the program types Grover (left) and QFT (right). 
    The lineplots compare Bloq and Proq with the ideal and noisy backends.}
    \label{fig:rq1_thresholds}
\end{figure}
\begin{table}
\caption{
Overall mean F1 scores per program type with bootstrapped \SI{99}{\percent} \gls{ci} and statistical comparison against Proq. 
\textbf{Threshold} is the index closest to the overall mean F1 score. 
\textbf{Mean F1} and \textbf{CI Half-Width} report the mean and its confidence interval. 
\textbf{MWU} indicates which approach performs better based on the Mann--Whitney U test: (Bloq), (Proq), or In-Sig. 
\textbf{A12} is the Vargha--Delaney effect size, and \textbf{Magnitude} classifies its strength as (N)egligible, (S)mall, (M)oderate, or (L)arge.
}
\label{tab:rq1_calculations}
\begin{tabular}{lllrrrccc}
\toprule
Program Type & Backend & Approach & \textbf{Threshold} & \textbf{Mean F1} & \textbf{CI Half-Width} & MWU & $\hat{A}_{12}$ & Magnitude \\
\midrule
Grover & Ideal & Bloq & 6 \% & 0.736079 & 0.008 & (Bloq) & 0.846923 & (L) \\
 &  & Proq & 14 \% & 0.380139 & 0.008 &  &  &  \\
 & Noisy & Bloq & 16 \% & 0.429974 & 0.011 & (Bloq) & 0.846923 & (L) \\
 &  & Proq & 17 \% & 0.056348 & 0.005 &  &  &  \\
\gls{qft} & Ideal & Bloq & 14 \% & 0.545863 & 0.001 & In-Sig & 0.515222 & (N) \\
 &  & Proq & 8 \% & 0.549616 & 0.001 &  &  &  \\
 & Noisy & Bloq & 16 \% & 0.441313 & 0.001 & In-Sig & 0.515222 & (N) \\
 &  & Proq & 13 \% & 0.426891 & 0.001 &  &  &  \\
\bottomrule
\end{tabular}
\end{table}

In \cref{fig:rq1_thresholds}, we show two plots, one for each program type, \gls{qft} to the right and Grover to the left.
Each plot shows the mean F1 score across all experiments for the corresponding threshold on the x-axis provided in percentages over the range of \SI{0}{\percent} to \SI{25}{\percent}.
In addition, we divide the overlaying lineplots by approach and backend with Bloq in green and Proq in orange, where the solid style depicts the ideal backend and dotted style depicts the noisy backend.
The shaded regions show the bootstrapped \SI{99}{\percent}  \gls{ci} around the mean values.

In addition, \cref{tab:rq1_calculations} shows two key values to assess the performance of each approach computed from the data in \cref{fig:rq1_thresholds}. 
First, the table shows the mean F1 scores across all experiments and all thresholds for a given approach, program type, and backend in the \textbf{Mean F1} column.
Second, in the \textbf{\SI{99}{\percent} CI} column, the table depicts the bootstrapped \SI{99}{\percent} \gls{ci} of the mean F1 score.

We now present and analyze the results from \cref{fig:rq1_thresholds} and \cref{tab:rq1_calculations} in two parts: 
\begin{inparaenum} 
    \item the ideal backend (solid lines) and 
    \item the noisy backend (dotted lines). 
\end{inparaenum}

\subsubsection{Ideal Backend}

\paragraph{\textbf{Grover}}
From the Grover program type in \cref{fig:rq1_thresholds}, we observe that at the \SI{0}{\percent} threshold, Bloq achieves a mean F1 score of \num{0.010984}, compared to Proq's \num{0.130935}.
This is expected, as Bloq requires a small baseline threshold due to its reliance on sampling to estimate expectation values for computing Bloch vectors.
For all the threshold values greater than \SI{0}{\percent}, Bloq's mean F1 score increases to \num{0.637681} as early as \SI{1}{\percent}, then follows a logarithmic trend, reaching its maximum mean score of \num{0.783921} at \SI{12}{\percent}.
We identify a stable threshold range between \SI{7}{\percent} and \SI{18}{\percent}, where Bloq's F1 score remains above \num{0.78}, indicating consistent high performance.
While \cref{fig:rq1_thresholds} shows mean F1 scores for a given threshold, from \cref{tab:rq1_calculations}, we observe that Bloq achieves a mean F1 score across all experiments and thresholds of \num{0.736079} with a \gls{ci} half-width of \num{0.009}.
From \cref{tab:rq1_calculations}. 
Furthermore, we observe that the closest threshold to this F1 score is \SI{6}{\percent}.

For Proq, beyond the \SI{0}{\percent} threshold, the mean F1 score increases linearly, reaching its maximum value of \num{0.564223} at \SI{25}{\percent}.
Turning to \cref{fig:rq1_thresholds}, which displays mean F1 scores as a function of the threshold, we observe that Bloq reaches its maximum mean F1 score at a lower threshold than Proq, implying that Bloq obtains a higher test precision with a lower uncertainty.  
From \cref{tab:rq1_calculations}, Proq achieves an overall mean F1 score of \num{0.380139} with a \gls{ci} half-width of \num{0.007}, with the closest threshold at \SI{14}{\percent}, implying again a higher uncertainty due to noise in the result.
Notably, even when averaging across all qubit counts, circuit depths, fault categories, and fault types, Bloq significantly outperforms Proq (\textbf{Bloq}, $\Tilde{A}_{12}$ = \num{0.846923}, \textbf{(L)}), indicating that the advantage is largely consistent throughout the experimental parameters considered.

\paragraph{\textbf{\gls{qft}}}  
Similar to Grover, Proq outperforms Bloq at the \SI{0}{\percent} threshold, which we attribute to Bloq's sensitivity to tiny thresholds, around \SI{1}{\percent}, due to its reliance on expectation value sampling.
At \SI{0}{\percent}, Proq achieves a maximum mean F1 score of \num{0.449996}, while Bloq reaches \num{0.355459}.
Beyond this threshold, both approaches follow a similar trend, with overall mean F1 scores of \num{0.545863} for Bloq and \num{0.549616} for Proq, and confidence interval half-widths on the order of \num{1e-4}, indicating high stability.
Both approaches reach their highest mean F1 score of \num{0.599541} at \SI{25}{\percent}.
However, the \gls{mwu} test indicates no statistically significant difference (\textbf{In-Sig}, $\Tilde{A}_{12}$ = \num{0.515222}), suggesting that when F1 scores are averaged across all qubit counts, circuit depths, fault categories, and fault types, the overall performances are statistically indistinguishable—though this may obscure differences at higher qubit counts or circuit depths.

\subsubsection{Noisy Backend}

\paragraph{\textbf{Grover}}  

Considering the noisy backend in \cref{fig:rq1_thresholds}, both approaches yield a mean F1 score of \num{0} at the \SI{0}{\percent} threshold.  
This is expected, as a zero-tolerance threshold inherently fails all test assessments due to noise.
Beyond the \SI{0}{\percent} threshold, Bloq follows a logarithmic increase, while Proq exhibits a linear trend, similar to the ideal backend.  
Bloq achieves an overall mean F1 score of \num{0.429974} at \SI{16}{\percent}, whereas Proq reaches \num{0.056348} at \SI{17}{\percent}.

Notably, Bloq's overall mean F1 score for the noisy backend surpasses Proq's with the ideal backend, with a \num{5} \gls{pp} improvement in favor of Bloq.
We attribute this improvement to Bloq's assertions, introducing a significantly lower depth overhead during debugging.
Comparing Bloq and Proq within the noisy case, Bloq outperforms Proq by \num{37} \gls{pp}.
Consistent with our results for the ideal backend, we observe from \cref{tab:rq1_calculations} that Bloq significantly outperforms Proq under noise as well (\textbf{Bloq}, $\Tilde{A}_{12}$ = \num{0.846923}, \textbf{(L)}).

\paragraph{\textbf{\gls{qft}}}  

For \gls{qft}, both approaches follow a similar trend at the \SI{0}{\percent} threshold, achieving a mean F1 score of \num{0.339310}.  
Beyond this threshold, they begin to deviate from the ideal backend.  
Bloq achieves an overall mean F1 score of \num{0.441313}, surpassing Proq by \num{1.4} \gls{pp} (\num{0.426891}).  
The largest deviation occurs at \SI{18}{\percent}, where Bloq outperforms Proq with a \num{2.8} \gls{pp} higher mean F1 score.
As in the ideal case, the approaches remain statistically indistinguishable under noise (\textbf{In-Sig}, $\Tilde{A}_{12}$ = \num{0.515222}), despite Bloq's slight advantage in mean F1 score.

\summarybox{RQ1 Summary}{
Bloq consistently outperforms Proq on Grover, with statistically significant and large effect sizes, especially under noise, where Bloq even surpasses Proq's ideal backend performance.
For \gls{qft}, both approaches perform similarly on both the noisy and ideal backend, with no statistically significant differences observed.
}

\subsection{RQ2: Localization by Qubit}

\begin{table}
\caption{F1 scores in the \textbf{Bloq} and \textbf{Proq} columns, Vargha-Delaney effect size ($\hat{\textbf{A}}_{12}$), statistical significance (\textbf{\gls{mwu}}), and effect size magnitude (\textbf{M}) for Bloq and Proq across different qubit counts and program types (Grover and \gls{qft}). The table presents results for both ideal and noisy backends, with statistical significance marked as \textbf{In-Sig} (insignificant), and the statistically favorable approach indicated as (Bloq) or (Proq) where applicable.}
\label{tab:rq2_stats}
\begin{tabular}{lrrrrlllrrrrll}
\toprule
 & & \multicolumn{5}{c}{Ideal Backend} & \multicolumn{5}{c}{Noisy Backend} \\
 \cmidrule(l{1pt}r{1pt}){3-7}
 \cmidrule(l{1pt}r{1pt}){8-12}
Program Type & Qubits & \textbf{Bloq} & \textbf{Proq} & $\hat{\textbf{A}}_{12}$ & \textbf{MWU} & \textbf{M} & \textbf{Bloq} & \textbf{Proq} & $\hat{\textbf{A}}_{12}$ & \textbf{MWU} & \textbf{M} \\
\midrule
\multirow[t]{5}{*}{Grover} & 2 & 0.3448 & 0.5455 & 0.1528 & (Proq) & (L) & 0.3448 & 0.5455 & 0.1528 & (Proq) & (L) \\
 & 3 & 0.5528 & 0.3304 & 0.8481 & (Bloq) & (L) & 0.4882 & 0.0000 & 1.0000 & (Bloq) & (L) \\
 & 4 & 0.6759 & 0.3059 & 0.9696 & (Bloq) & (L) & 0.4667 & 0.0000 & 1.0000 & (Bloq) & (L) \\
 & 5 & 0.7271 & 0.1524 & 1.0000 & (Bloq) & (L) & 0.3192 & 0.0000 & 1.0000 & (Bloq) & (L) \\
 & 6 & 0.8538 & 0.2061 & 1.0000 & (Bloq) & (L) & 0.1781 & 0.0000 & 0.9714 & (Bloq) & (L) \\
\cline{1-12}
\multirow[t]{9}{*}{\gls{qft}} & 2 & 0.7814 & 0.7814 & 0.5000 & In-Sig & (N) & 0.7814 & 0.7814 & 0.5000 & In-Sig & (N) \\
 & 3 & 0.7037 & 0.7037 & 0.5000 & In-Sig & (N) & 0.7037 & 0.7037 & 0.5000 & In-Sig & (N) \\
 & 4 & 0.6430 & 0.6433 & 0.4980 & In-Sig & (N) & 0.6402 & 0.6402 & 0.4993 & In-Sig & (N) \\
 & 5 & 0.6096 & 0.6096 & 0.5000 & In-Sig & (N) & 0.6026 & 0.6015 & 0.5019 & In-Sig & (N) \\
 & 6 & 0.5751 & 0.5751 & 0.5000 & In-Sig & (N) & 0.5603 & 0.5588 & 0.5066 & In-Sig & (N) \\
 & 7 & 0.5522 & 0.5522 & 0.5000 & In-Sig & (N) & 0.4334 & 0.4322 & 0.5043 & In-Sig & (N) \\
 & 8 & 0.5372 & 0.5372 & 0.5000 & In-Sig & (N) & 0.3273 & 0.3269 & 0.5030 & In-Sig & (N) \\
 & 9 & 0.5229 & 0.5229 & 0.5000 & In-Sig & (N) & 0.3367 & 0.3305 & 0.5432 & In-Sig & (N) \\
 & 10 & 0.5116 & 0.5116 & 0.4999 & In-Sig & (N) & 0.3432 & 0.3176 & 0.6611 & (Bloq) & (S) \\
\cline{1-12}
\bottomrule
\end{tabular}
\end{table}

In \cref{tab:rq2_stats}, we present the results by qubit count, assessing the relationship between qubit count and the effectiveness of each approach.
We present the mean F1 scores for each approach in the columns \textbf{Bloq} and \textbf{Proq}.  
The results of our statistical tests, conducted using the \gls{mwu} test, are shown in the \textbf{\gls{mwu}} column.  
Here, \textbf{In-Sig} indicates an insignificant difference between F1 scores, while the name of the statistically favorable approach (Proq or Bloq) is displayed otherwise.  
Additionally, the Vargha-Delaney effect size $\hat{A}_{12}$ is reported in the \textbf{A12} column, with values ranging from 0 to 1, alongside the corresponding effect size magnitude category in the \textbf{M} column.

\paragraph{\textbf{Grover}}  
From \cref{tab:rq2_stats}, Proq shows a large statistically significant advantage at 2 qubits in both the ideal and noisy backend.  
However, for all the cases beyond 2 qubits, Bloq significantly outperforms Proq.

We now examine the trend of F1 scores as the number of qubits increases.  
At 2 qubits, Bloq starts with an F1 score of \num{0.3448}, outperformed by Proq's \num{0.5455}.  
Beyond this, Bloq's F1 score steadily increases with each additional qubit, reaching \num{0.8538}.
Compared to Proq, the F1 score drops to \num{0.3304} at 3 qubits, decreases further to \num{0.1524}, and then rises to \num{0.2061} at 6 qubits.  
This results in a \SI{0.64769}{\percent} point improvement when Bloq is applied at 6 qubits.  

For the noisy backend, both approaches are affected by noise, with Proq's effectiveness dropping to zero beyond 2 qubits.  
Bloq demonstrates greater robustness, maintaining an F1 score of \num{0.4667} at 4 qubits before decreasing to \num{0.1781} at 6 qubits as noise increasingly impacts the results.

\paragraph{\textbf{\gls{qft}}}  
From \cref{tab:rq2_stats}, we observe no statistically significant differences between Proq and Bloq across all qubit counts, except at 10 qubits when applying the noisy backend.  
At this highest qubit count---corresponding to the deepest circuits---Bloq is significantly more effective than Proq with a small effect size.  
In this case, Bloq achieves an F1 score of \num{0.3432}, compared to Proq's \num{0.3176}.

\begin{summarybox}{RQ2 Summary}{
For Grover, Bloq scales better than Proq, outperforming it beyond 2 qubits and retaining high F1 scores even under noise, where Proq's performance degrades sharply.
For \gls{qft}, the approaches perform similarly at low qubit counts, but as qubit counts and circuit depths increase, Bloq statistically outperforms Proq.
}
\end{summarybox}

\subsection{RQ3: Localization by Fault Segment}  
\label{sec:results_rq3}

\begin{table}
\caption{F1 scores for Bloq and Proq across different fault segments (0 to 5 for Grover and 0 to 9 for \gls{qft}) under ideal and noisy backends. The table also includes the Vargha-Delaney effect size ($\hat{\textbf{A}}_{12}$), statistical significance (\textbf{\gls{mwu}}), and effect size magnitude (\textbf{M}). Statistical significance is marked as \textbf{In-Sig} (insignificant), and the statistically favorable approach is indicated as (Bloq) or (Proq) where applicable.}
\label{tab:rq4_stats}
\begin{tabular}{lrrrrlllrrrrll}
\toprule
 & & \multicolumn{5}{c}{Ideal Backend} & \multicolumn{5}{c}{Noisy Backend} \\
 \cmidrule(l{1pt}r{1pt}){3-7}
 \cmidrule(l{1pt}r{1pt}){8-12}
Program Type & Faulty Segment & \textbf{Bloq} & \textbf{Proq} & $\hat{\textbf{A}}_{12}$ & \textbf{\gls{mwu}} & \textbf{M} & \textbf{Bloq} & \textbf{Proq} & $\hat{\textbf{A}}_{12}$ & \textbf{\gls{mwu}} & \textbf{M} \\
\midrule
\multirow[t]{6}{*}{Grover} & 1 & 0.8099 & 0.3642 & 0.9617 & (Bloq) & (L) & 0.5535 & 0.0145 & 0.9835 & (Bloq) & (L) \\
 & 2 & 0.8102 & 0.1119 & 0.9991 & (Bloq) & (L) & 0.3775 & 0.0000 & 0.9583 & (Bloq) & (L) \\
 & 3 & 0.8082 & 0.1141 & 0.9971 & (Bloq) & (L) & 0.1327 & 0.0000 & 0.8482 & (Bloq) & (L) \\
 & 4 & 0.7912 & 0.1339 & 1.0000 & (Bloq) & (L) & 0.0954 & 0.0000 & 0.7135 & (Bloq) & (M) \\
 & 5 & 0.7816 & 0.1815 & 0.9998 & (Bloq) & (L) & 0.0210 & 0.0000 & 0.5547 & In-Sig & (N) \\
 & 6 & 0.8221 & 0.3424 & 1.0000 & (Bloq) & (L) & 0.1509 & 0.0000 & 0.9453 & (Bloq) & (L) \\
\cline{1-12}
\multirow[t]{10}{*}{\gls{qft}} & 1 & 0.6998 & 0.6998 & 0.5000 & In-Sig & (N) & 0.6582 & 0.6549 & 0.4994 & In-Sig & (N) \\
 & 2 & 0.5908 & 0.5908 & 0.5000 & In-Sig & (N) & 0.6094 & 0.5692 & 0.5894 & (Bloq) & (S) \\
 & 3 & 0.5286 & 0.5286 & 0.5000 & In-Sig & (N) & 0.4832 & 0.4422 & 0.6275 & (Bloq) & (S) \\
 & 4 & 0.4791 & 0.4792 & 0.4999 & In-Sig & (N) & 0.3775 & 0.3587 & 0.5650 & In-Sig & (N) \\
 & 5 & 0.4633 & 0.4633 & 0.5000 & In-Sig & (N) & 0.2651 & 0.2687 & 0.4884 & In-Sig & (N) \\
 & 6 & 0.4478 & 0.4478 & 0.5000 & In-Sig & (N) & 0.2267 & 0.2302 & 0.4864 & In-Sig & (N) \\
 & 7 & 0.4571 & 0.4571 & 0.5000 & In-Sig & (N) & 0.2289 & 0.2267 & 0.5109 & In-Sig & (N) \\
 & 8 & 0.4836 & 0.4836 & 0.5000 & In-Sig & (N) & 0.2430 & 0.2304 & 0.5585 & In-Sig & (N) \\
 & 9 & 0.4925 & 0.4925 & 0.4999 & In-Sig & (N) & 0.2391 & 0.2231 & 0.5401 & In-Sig & (N) \\
 & 10 & 0.6427 & 0.6427 & 0.5000 & In-Sig & (N) & 0.4454 & 0.3982 & 0.6636 & (Bloq) & (S) \\
\cline{1-12}
\bottomrule
\end{tabular}
\end{table}

In \cref{tab:rq4_stats}, we categorize F1 scores based on the program segment where the fault was inserted.  

\paragraph{\textbf{Grover}}  
For the ideal backend, Bloq achieves a statistically significant improvement over Proq, maintaining an F1 score of at least \num{0.7816} across all the segments.  
This indicates that Bloq's effectiveness remains stable regardless of circuit depth.  
Proq attains its highest F1 scores in the first and last segments (\num{0.3642} and \num{0.3424}, respectively), while for all the other segments, its F1 score ranges between \num{0.1119} and \num{0.1815}.
With the noisy backend, both approaches show reduced effectiveness in later segments, but Proq is particularly impacted.  
It achieves a minimal F1 score of \num{0.0145} in the first segment and has zero fault detection effectiveness across all the other segments.
Comparing Bloq’s F1 score in the first segment, it achieves \num{0.5535}, maintaining effectiveness.  
However, its effectiveness gradually declines with circuit depth, remaining viable in the second and third segments.  
By the fifth segment, there is no significant difference between Bloq and Proq, indicating a loss of effectiveness due to noise.
Despite this, Bloq's F1 score resurfaces at \num{0.1509} in the sixth segment, indicating that a loss of effectiveness in earlier segments does not preclude fault detection in later segments (see \cref{sec:discuss_segment_effects} for discussion).

\paragraph{\textbf{\gls{qft}}}  

As in the previous research questions, when applying the ideal backend, \gls{qft} shows no significant differences between Bloq and Proq.  
Both approaches achieve an F1 score of \num{0.6998} in the first segment, with effectiveness gradually decreasing to a minimum of \num{0.4478} in the sixth segment before increasing again to \num{0.6427} at the tenth segment.  
This trend suggests that both approaches are less effective in the middle segments but regain effectiveness toward the later segments, exhibiting a symmetric pattern.
We attribute this to the fact that the center of the circuit for both Grover and \gls{qft} have a higher degree of entanglement~\citep{QCQIBook,GroversAlgo}, increasing the difficulty of fault localization (see \cref{sec:discuss_segment_effects} for discussion).

For the noisy backend, Bloq outperforms Proq with a small but significant effect size in the second, third, and tenth segments, while no significant differences are observed in other segments.  
Specifically, Bloq achieves a \num{4} \gls{pp} improvement over Proq in the second segment, \num{4.1} \gls{pp} in the third, and \num{4.7} \gls{pp} in the tenth.

Building on the symmetry observed with the ideal backend, we now examine the general F1 score trend for the noisy backend.  
Both Bloq and Proq start at \num{0.65} before gradually decreasing.  
While the ideal backend exhibited a minimum F1 score followed by an increase, no clear symmetric minimum is observed for the noisy backend.
The minimum F1 score for the noisy backend occurs at the ninth segment with a value of \num{0.2231}, with adjacent segments showing similar values, such as \num{0.2289} for Bloq in the seventh segment.  
However, as for the ideal backend, we observe a significant increase at the tenth segment, where Bloq achieves a statistically significant improved F1 score of \num{0.4454} compared to Proq’s \num{0.3982}.

\begin{summarybox}{RQ3 Summary}{
Bloq outperforms Proq across all circuit segments for Grover under ideal conditions and retains high effectiveness under noise, where Proq degrades rapidly with circuit depth.  
For \gls{qft}, both approaches perform similarly overall for the ideal backend, though Bloq shows significant improvements in early and late segments for the noisy backend.
}
\end{summarybox}

\subsection{RQ4: Localization by Fault Category}

\begin{table}
\caption{F1 scores for Bloq and Proq across different fault categories (\addgate{}, \removegate{}, \replacegate{}) for Grover and \gls{qft} programs. 
The table shows results for both ideal and noisy backends, including Vargha-Delaney effect size (\textbf{A12}), statistical significance (\textbf{\gls{mwu}}), and effect size magnitude (\textbf{M}). Statistical significance is indicated as \textbf{In-Sig} (insignificant), and the statistically favorable approach is noted as (Bloq) or (Proq) where applicable.}
\label{tab:rq3_stats}
\begin{tabular}{llrrrlllrrrrll}
\toprule
 & & \multicolumn{5}{c}{Ideal Backend} & \multicolumn{5}{c}{Noisy Backend} \\
 \cmidrule(l{1pt}r{1pt}){3-7}
 \cmidrule(l{1pt}r{1pt}){8-12}
Program Type & Fault Category & \textbf{Bloq} & \textbf{Proq} & $\hat{\textbf{A}}_{12}$ & \textbf{MWU} & \textbf{M} & \textbf{Bloq} & \textbf{Proq} & $\hat{\textbf{A}}_{12}$ & \textbf{\gls{mwu}} & \textbf{M} \\
\midrule
\multirow[t]{3}{*}{Grover} & \addgate{} & 0.7881 & 0.2213 & 0.9906 & (Bloq) & (L) & 0.2186 & 0.0017 & 0.9576 & (Bloq) & (L) \\
 & \removegate{} & 0.8029 & 0.2398 & 0.9804 & (Bloq) & (L) & 0.2315 & 0.0020 & 0.9451 & (Bloq) & (L) \\
 & \replacegate{} & 0.8231 & 0.1670 & 0.9905 & (Bloq) & (L) & 0.2159 & 0.0013 & 0.9655 & (Bloq) & (L) \\
\cline{1-12}
\multirow[t]{3}{*}{\gls{qft}} & \addgate{} & 0.5299 & 0.5300 & 0.4999 & In-Sig & (N) & 0.3635 & 0.3431 & 0.6366 & (Bloq) & (S) \\
 & \removegate{} & 0.4399 & 0.4399 & 0.5000 & In-Sig & (N) & 0.2947 & 0.2913 & 0.5176 & In-Sig & (N) \\
 & \replacegate{} & 0.5688 & 0.5688 & 0.5000 & In-Sig & (N) & 0.3756 & 0.3503 & 0.6555 & (Bloq) & (S) \\
\bottomrule
\end{tabular}
\end{table}

In \cref{tab:rq3_stats}, we categorize the F1 scores based on fault type: \addgate{}, \replacegate{}, and \removegate{} gates, as defined in \cref{sec:eval_design}.

\paragraph{\textbf{Grover}}  
Similar to \cref{tab:rq1_calculations}, Bloq achieves a statistically significant improvement over Proq across all the fault categories in both ideal and noisy backends.  
For the ideal backend, Bloq attains F1 scores of at least \num{0.7881} across all the categories, whereas Proq's F1 scores range from \num{0.1670} to \num{0.2213}, highlighting Bloq's substantial advantage.
With the noisy backend, Bloq's F1 scores drop to between \num{0.2159} and \num{0.2315}, indicating a significant reduction in effectiveness due to noise.  
For Proq, the F1 scores decline even further, reaching the order of magnitude of \num{e-3}.

\paragraph{\textbf{\gls{qft}}}  
For ideal backend, most statistical tests show no significant difference between Bloq and Proq across all the fault categories.  
Both approaches achieve F1 scores ranging from \num{0.43} to \num{0.57}.
For the noisy backend, Bloq shows a small but statistically significant advantage over Proq for the add and replace fault categories, while no significant difference is observed for remove.  
Bloq achieves an F1 score of \num{0.3635} for add, compared to Proq's \num{0.3431}, and \num{0.3756} for replace, compared to Proq's \num{0.3503}.  
Thus, applying Bloq in the presence of noise results in a \num{2} \gls{pp} improvement for add faults and \num{2.5} \gls{pp} \SI{2.5}{\percent} for replace faults.

\begin{summarybox}{RQ4 Summary}{
Bloq consistently outperforms Proq across all the fault categories for Grover, with especially strong results for the ideal backend and considerably higher performance under noise.
For \gls{qft}, both approaches perform similarly, though Bloq achieves a small but significant improvement for \addgate{} and \replacegate{} faults under noise.
}
\end{summarybox}

\subsection{RQ5: Runtime and Assertion Depth Overhead}  

\begin{table}
\caption{Summary of test runtime statistics for Bloq and Proq across different program types (Grover and \gls{qft}) and backends (ideal and noisy). The table includes the mean, median, and interquartile range (\gls{iqr}) of runtime for each approach, highlighting the differences in computational efficiency.}
\label{tab:rq5_runtime_stats}
\begin{tabular}{lllrrr}
\toprule
        \multicolumn{3}{c}{} & \multicolumn{3}{c}{Test Runtime Statistics} \\
 \cmidrule(l{1pt}r{1pt}){4-6}
Program Type & Backend & Approach & \textbf{Mean} & \textbf{Median} & \textbf{IQR} \\
\midrule
\multirow[t]{4}{*}{Grover} & \multirow[t]{2}{*}{Ideal} & Bloq & 1.31 & 1.27 & 0.25  \\
 &  & Proq & 6.05 & 8.28 & 6.50  \\
\cline{2-6}
 & \multirow[t]{2}{*}{Noisy} & Bloq & 28.91 & 28.76 & 5.55  \\
 &  & Proq & 143.61 & 201.19 & 165.07 \\
\cline{1-6} \cline{2-6}
\multirow[t]{4}{*}{\gls{qft}} & \multirow[t]{2}{*}{Ideal} & Bloq & 0.24 & 0.22 & 0.04  \\
 &  & Proq & 0.04 & 0.04 & 0.01  \\
\cline{2-6}
 & \multirow[t]{2}{*}{Noisy} & Bloq & 5.19 & 5.05 & 0.71  \\
 &  & Proq & 4.52 & 4.60 & 1.00  \\
\cline{1-6} \cline{2-6}
\bottomrule
\end{tabular}
\end{table}
  
\begin{table}
\caption{Summary of circuit test depth statistics for Bloq and Proq across different program types (Grover and \gls{qft}). The table includes the mean, median, interquartile range (\gls{iqr}), and quartiles (\gls{q1}, \gls{q3}) for each approach, highlighting the differences in circuit depth overhead between the approaches.}
\label{tab:rq5_depth_stats}
\begin{tabular}{llrrrrr}
\toprule
        \multicolumn{2}{c}{} & \multicolumn{5}{c}{Depth Overhead Statistics} \\
 \cmidrule(l{1pt}r{1pt}){3-7}
Program Type & Approach & \textbf{Mean} & \textbf{Median} & \textbf{\gls{iqr}} & \textbf{\gls{q1}} & \textbf{\gls{q3}} \\
\midrule
\multirow[t]{2}{*}{Grover} & Bloq & 229.85 & 268.00 & 122.00 & 147.00 & 269.00 \\
 & Proq & 5287.51 & 7189.00 & 5194.00 & 2049.00 & 7243.00 \\
\cline{1-7}
\multirow[t]{2}{*}{\gls{qft}} & Bloq & 7.11 & 7.00 & 0.00 & 7.00 & 7.00 \\
 & Proq & 131.65 & 134.00 & 34.00 & 116.00 & 150.00 \\
\cline{1-7}
\bottomrule
\end{tabular}
\end{table}

In \cref{tab:rq5_runtime_stats,tab:rq5_depth_stats}, we present the test runtime and circuit depth requirements for both approaches, categorized by program type and backend.  
The reported results include the mean, median, and \gls{iqr} for both runtime and depth, computed across all results.
Because the depth overhead results contain outliers due to the variability in the qubit count across experiments, we employ the median along with \gls{iqr} and quartiles to quantify the spread.
In addition, we include the mean for comparison.

\subsubsection{Runtime}

Overall, Bloq exhibits significantly lower runtime compared to Proq, particularly for Grover, where it is consistently faster and has lower variability. 
However, in some cases, Proq achieves shorter runtimes, notably for \gls{qft} for the ideal backend.

For Grover with the ideal backend, Bloq achieves a significantly faster mean runtime of \SI{1.31}{\second} compared to Proq's \SI{6.05}{\second}, with a median runtime of \SI{1.27}{\second} versus \SI{8.28}{\second}. 
The variability of Proq is also significantly higher, as reflected in its \gls{iqr} of \SI{6.50}{\second} compared to Bloq's \SI{0.25}{\second}. 
For the noisy backend, Bloq maintains an advantage with a mean runtime of \SI{28.91}{\second}, while Proq experiences a significant increase to \SI{143.61}{\second}, with a \gls{iqr} of \SI{165.07}{\second}, indicating large runtime fluctuations.
Additionally, for Grover with the noisy backend, the median runtime of Proq is close to \gls{q3} (\SI{201.19}{\second}), suggesting that a significant portion of the observed results are skewed toward slower runtimes.

For \gls{qft}, the trends are different. 
With the ideal backend, Proq is faster, with a mean runtime of \SI{0.04}{\second} compared to Bloq's \SI{0.24}{\second}, and lower variability. 
With the noisy backend, Bloq is slower than Proq on average (\SI{5.19}{\second} vs. \SI{4.52}{\second}), but Proq exhibits greater variability, with an \gls{iqr} of \SI{1.00}{\second} compared to Bloq’s \SI{0.71}{\second}.

These findings indicate that Bloq is generally the more computationally efficient approach, particularly for Grover, where it achieves significantly lower runtimes and less variability across both the ideal and noisy backends. 
However, for \gls{qft}, Proq demonstrates advantages in execution speed for the ideal backend and remains competitive under noise.

\subsubsection{Assertion Depth Overhead}

In \cref{tab:rq5_depth_stats}, we quantify the depth overhead introduced by each approach across different program types.
This depth accounts for all circuit overhead required to reach a test result corresponding to the test runtimes of \cref{tab:rq5_runtime_stats}.

For Grover, Bloq consistently results in lower-depth circuits, achieving a mean depth of \num{229.85}, whereas Proq introduces a significantly larger depth overhead, with a mean depth of \num{5287.51}. 
The variability in Proq's depth is also very high, with an \gls{iqr} of \num{5194.00}, suggesting a wide spread in circuit complexity across different test cases.
Furthermore, Proq's \gls{q1} value of \num{2049.00} and \gls{q3} of \num{7243.00} indicate that while some circuits have moderate circuit depths, many cases exhibit large circuit depths.
Bloq also exhibits a high variability in circuit depth for Grover, with an \gls{iqr} of \num{122.00}. 
However, these large variances is mostly attributed to the statistics in \cref{tab:rq5_depth_stats} being calculated across all the qubit counts, meaning that a variability should be expected due to different qubit counts.

For \gls{qft}, the trends differ. 
Bloq maintains consistently low circuit depths across all cases, with a mean depth of \num{7.11} and an \gls{iqr} of \num{0}, indicating minimal variability. 
Proq, on the other hand, has a larger mean depth of \num{131.65}, though its variability is lower than in the Grover case, with an \gls{iqr} of \num{34.00}. 

These results indicate that Bloq consistently introduces less depth overhead than Proq, particularly for Grover, where the depth difference is substantial.
In \gls{qft}, Proq has a moderate increase in depth but remains within a reasonable range, whereas Bloq maintains minimal depth overhead.

\begin{summarybox}{RQ5 Summary}{
Bloq significantly reduces both runtime and circuit depth compared to Proq for Grover, especially under noise, where it is considerably faster and has lower variability.  
For \gls{qft}, Proq is faster, but Bloq achieves significantly lower assertion depth overhead and reduced variability under noise.
}
\end{summarybox}

\section{Discussion}
\label{sec:discussion}

This section discusses and reflects on the key findings of our evaluation.
We structure the discussion as follows: 
\begin{inparaenum}[(1)]
    \item we provide reflections and interpretations of the experimental results presented in \cref{sec:results}, structured around our research questions RQ1--RQ5;
    \item then, we discuss how Bloq addresses the specific challenges of runtime assertion testing identified in \cref{sec:introduction};
    \item here, we discuss how Bloq may offer additional capabilities beyond fault localization, such as fault identification or program repair;
    and
    \item finally, we outline the limitations of our work and discuss avenues for future research.
\end{inparaenum}

\subsection{Interpretations of Results}
\label{sec:discuss_interpret_results}

We now provide interpretations of our findings based on the results presented in \cref{sec:results}, structured around our research questions (RQ1--RQ5).

\subsubsection{Circuit Depth Drives Localization Effectiveness}

Across RQ1--RQ4, our results show that Bloq consistently outperforms Proq in fault localization effectiveness for both Grover and \gls{qft}.
However, the magnitude of this advantage is closely tied to the structure and depth of the quantum program under test.

For Grover, which is characterized by a large increasing depth with increasing qubit count, Bloq shows substantial improvements over Proq across all the evaluated settings for both effectiveness and efficiency. 
These benefits persist regardless of circuit segment (RQ3) or fault category (RQ4), and become more pronounced as the number of qubits increases (RQ2).

For \gls{qft}, which has significantly lower circuit depths than Grover, Bloq performs comparably to Proq at lower qubit counts. 
However, as the number of qubits increases to 10, Bloq begins to demonstrate measurable advantages in fault detection effectiveness, with statistically significant improvements over Proq under noise (RQ2, RQ3, RQ4).

Together, these findings demonstrate that Bloq's advantages are maximized in quantum programs exhibiting substantial depth and noise, while still offering comparable performance in lower-depth and less noisy circuits.

\subsubsection{Higher Effectiveness in Early and Late Segments}
\label{sec:discuss_segment_effects}

From our RQ3 results in \cref{tab:rq3_stats}, we observe increased fault localization effectiveness in the early and late circuit segments across both Grover and \gls{qft}. 
While the increased effectiveness in early segments can be attributed to lower noise levels and low assertion depth, the elevated effectiveness in the later segments of Proq for Grover and \gls{qft}, and of Bloq for \gls{qft}, warrants further discussion.
For Grover, Proq uses multiples of the Grover operator in its assertions, which increase in depth in later segments. 
Thus, when a fault is present, the added depth in later Proq assertions increases fault propagation, making detection more likely across affected qubits.

In \gls{qft}, both Bloq and Proq show reduced F1 scores for the middle segments in \cref{tab:rq3_stats}, while early and late segments achieve higher scores. 
In this case, a segment corresponds to a single qubit, so middle segments refer to the more central qubits in the \gls{qft} circuit. 
These qubits participate more heavily in entangling operations, acting as both control and target in controlled rotations.
Due to our segment definition strategy, where each segment consists of a single qubit, faults may propagate through entanglement and trigger failures in assertions on non-faulty qubits.
These failures will then not be detectable by our test assessment due to the segment definition.
This results in false positives and false negatives—cases which should instead be registered as true positives. 
This limitation affects the fault detection effectiveness of both Proq and Bloq equally in \gls{qft}.
In future work, this could be addressed by evaluating all qubits per segment, as we did for Grover, to improve detection of entanglement-driven fault propagation.

\subsubsection{Scalability and Robustness under Noise}

Our results across RQ2, RQ3, and RQ5 demonstrate that Bloq scales better than Proq with increasing qubit count and circuit depth, particularly for the noisy backend.
For Grover, Bloq's effectiveness improves with increasing qubit count, achieving F1 scores as high as \num{0.85} at \num{6} qubits for the ideal backend (RQ2).
In contrast, Proq's performance degrades drastically beyond \num{2} qubits and fails entirely under noise for higher qubit counts and depths.
We observe a similar trend across circuit segments (RQ3), where Bloq maintains its effectiveness even in segments at high depth, while Proq becomes ineffective beyond the first segment in the presence of noise.

For \gls{qft}, while both approaches perform comparably at smaller qubit counts, Bloq becomes more effective than Proq under noise at \num{10} qubits and in later segments (RQ2, RQ3).
This suggests that Bloq yields increased resilience to noise as the depth and noise of the circuit increase.
We attribute this resilience to Bloq assertion's low overhead (RQ5), causing less noise to be introduced into the circuit.

In addition to effectiveness, Bloq exhibits superior efficiency for Grover by a factor of \num{5} for the mean runtime and a reduction in mean depth overhead by a factor of \num{23} compared to Proq (RQ5), directly addressing challenge (1) from \cref{sec:introduction}.
For \gls{qft}, Proq outperforms Bloq's mean runtime by a factor of \num{1.15} for the noisy backend, but at the cost of increased mean depth overhead by a factor of \num{19}.
We attribute Proq's better performance to \gls{qft}'s lower depth compared to Grover.
This is supported by Proq's higher median runtime in \cref{tab:rq5_runtime_stats} for the noisy backend compared to the mean runtime, indicating that the programs with higher depths and qubits are skewing the distribution towards higher runtimes.
This is because Proq inserts all assertions into the circuit and must execute all of them, regardless of where the fault occurs.  
Even when a fault is located in an early segment, the entire circuit, including all assertion overhead, must still be run, leading to significant runtime increases as circuit depth and qubit count grow.

Overall, these findings indicate that Bloq not only localizes faults more effectively as circuits grow larger with increased depths but also offers practical advantages in terms of runtime, depth overhead, and robustness to noise.
All of these are critical factors for the scalability of quantum software testing on \gls{nisq} devices.
These factors directly alleviate challenge (2) from \cref{sec:introduction}, which concerns the scalability of assertion-based testing on \gls{nisq} devices.

\subsubsection{Practical Advantages for \gls{nisq} Systems}

In addition to reduced circuit depth and runtime, Bloq offers several practical advantages for \gls{nisq} systems.  
Unlike Proq, which relies heavily on mid-circuit measurements, Bloq avoids this requirement.  
Mid-circuit measurements require advanced hardware capabilities, including fast, synchronized readouts and dynamic circuit control—features, which are not uniformly supported across quantum devices~\citep{ChenMidCircuit}.  
Moreover, these measurements are required in each Proq assertion, contributing to a mean circuit depth of \num{5287.51} for Grover, compared to only \num{229.85} for Bloq.  
By avoiding mid-circuit measurement overhead, Bloq reduces its vulnerability to measurement-induced noise and increases its compatibility with \gls{nisq} systems.

Moreover, Bloq enables terminating the test execution early when a fault is detected.  
In contrast, Proq must execute the full circuit, including all assertions, even when faults are located in early segments.  
This contributes to Proq’s substantially higher mean runtime under noise: \SI{143.61}{\second} for Grover, compared to Bloq's \SI{28.91}{\second}.  
Bloq's early-exit capability is particularly valuable on \gls{nisq} hardware, where extended runtimes increase exposure to circuit noise.

Finally, Bloq's assertion scheme is parallelizable, as each assertion operates independently.  
This enables distributing Bloq assertions across multiple quantum backends or executing them concurrently using backend-supported batching~\citep{QBatching_Khare}.  
Given that Bloq's total mean runtime for Grover is under \SI{30}{\second}, substantial speedups can be achieved by parallel execution, making Bloq well-suited for environments with limited quantum hardware availability.
Together, these features make Bloq a highly practical solution for assertion-based fault localization on \gls{nisq} devices, where noise resilience, reduced assertion overhead, and parallelism are critical for scalability.

\subsection{AutoBloq's Automation and Generalizability}
\label{sec:autobloq_generalizability}

Our evaluation confirms that AutoBloq enables fully automated insertion of Bloq assertions for both Grover and \gls{qft} without requiring any circuit-specific manual design.
This demonstrates that assertion automation for realistic quantum programs is achievable in practice, directly addressing challenge (3) from \cref{sec:introduction}.

As described in \cref{def:autobloq}, AutoBloq relies on algorithm specifications introduced in \cref{sec:background_quantum_algorithms}, where each segment defines an expected transformation on each qubit.  
This makes the approach applicable to any quantum algorithm that can be decomposed into a sequence of segment-wise operations.
Thus, we expect AutoBloq to generalize well to quantum algorithms with a modular and deterministic structure, such as quantum walks, quantum phase estimation, amplitude amplification, and modular exponentiation in Shor's algorithm~\citep{Venegas_Andraca_2012_Quantum_Walk,QCQIBook,Shor_1997}.  
These algorithms exhibit well-defined unitary operations applied in a repeatable pattern, which aligns with AutoBloq.  
Parameterized quantum algorithms like \gls{vqe} and \gls{qaoa} are not suited for AutoBloq, as each iterative step depends on expectation values measured in previous iterations~\citep{Moll_QAOA,QCQIBook}. 
This feedback loop prevents a fixed algorithm specification, which AutoBloq relies on to define expected qubit states.

To support such algorithms, the algorithm specification in \cref{sec:background_quantum_algorithms} could be extended to include parameter dependencies between iterations.  
This would enable assertion generation based on the parameters in the optimization loop.

\subsection{Need for Empirical Evaluations}

Finally, this work contributes to addressing the gap in empirical evaluations of runtime assertion techniques for quantum programs.
Our extensive experimental evaluation not only provides the first in-depth assessment of Bloq assertions but also offers, to our knowledge, the first empirical evaluation of Proq assertions, complementing the original case study presented by its authors.

Through our evaluation of \gls{qft}, we also provide an example of how projective assertion schemes can be automated for approaches such as Proq.
Notably, while Bloq outperforms Proq in most settings in our evaluation, our results show that Proq can still perform well in low-depth, less-noisy circuits.
In specific configurations, such as the zero-percent threshold in RQ1 for the ideal backend, Proq even demonstrates higher efficiency than Bloq.
However, this efficiency comes at the cost of lower effectiveness compared to what Bloq achieves at higher thresholds, where it consistently outperforms Proq.

We believe that this evaluation framework, spanning multiple fault categories, segments, noise levels, and program types, can serve as a template for future experimental studies of runtime assertion techniques in quantum software engineering.

\subsection{Beyond Fault Localization: Additional Benefits of Bloch Vector-Based Assertions}
\label{sec:discussion_additional}

Beyond enabling experimentally-efficient fault localization, our findings suggest that Bloq assertions may offer additional capabilities, including fault identification and even potential for program repair.
The core idea builds on the observation that, by measuring the Bloch vector of a single qubit at a given segment, and comparing it to the expected Bloch vector generated by AutoBloq, we effectively have access to both the faulty and expected density matrices for that qubit.
This opens the possibility of inferring a transformation---a so-called fault equivalent matrix \(F\)---that relates the faulty state to the expected state via:

\begin{equation}
\rho_{q,k} = F \sigma_{q,k} F^{\dagger},
\label{eq:density_classify_repair_transformation}
\end{equation}

Where in \cref{eq:density_classify_repair_transformation}, \(\rho_{q,k}\) denotes the measured (faulty) density matrix, and \(\sigma_{q,k}\) the expected (non-faulty) density matrix.
Accurate reconstruction of \(F\) requires that \(\rho\) and \(\sigma\) share the same eigenvalues, since unitary transformations preserve the spectrum of density matrices~\citep{QCQIBook}.
Given this relation, \(F\) provides valuable diagnostic information about the type of fault that has occurred. 
In principle, applying \(F^{\dagger}\) could be used to correct the fault, restoring the original expected state:
\[
F^{\dagger} \rho_{q,k} F = \sigma_{q,k}.
\]

\paragraph{Example: Identifying and Repairing a Fault}

Consider a simple scenario where the expected density matrix of a qubit is the pure state along the \(X\)-axis:

\begin{equation}
\sigma = \frac{1}{2} (\mathbbm{1} + X)
\end{equation}

However, suppose the measured faulty density matrix is:
\[
\rho = \frac{1}{2} (\mathbbm{1} - X)
\]
This indicates that the Bloch vector has flipped along the \(X\)-axis, suggesting the action of a \(Y\) gate fault (since \(Y X Y^{\dagger} = -X\)).

To confirm this, we can compute the fault equivalent matrix \(F\) that satisfies:
\[
\rho = F \sigma F^{\dagger}
\]
In this case, it is straightforward to verify that \(F = Y\) satisfies the equation. 
Therefore, we have not only localized the fault but also classified it as a \(Y\) gate.

To repair the circuit, one could insert an additional \(Y\) gate at the faulty location, such that:
\[
Y \rho Y^{\dagger} = \sigma
\]
restoring the expected density matrix.

This illustrates how Bloq assertions could, in certain scenarios, enable both fault identification and repair.
However, we note that this fault classification and repair capability relies on the assumption that the transformation between the expected and measured density matrices is unitary, which requires that the two matrices share the same eigenvalues~\citep{QCQIBook}. 
In the presence of noise, this condition may not hold, potentially reducing the accuracy of both fault classification and repair. Further studies are needed to evaluate the extent to which noise impacts these capabilities in practical settings.

This idea of program repair in quantum computing has been explored in various forms in prior work~\citep{AutomaticRepairQuantum, repairGPT}. 
However, existing approaches typically rely on expensive searches in unitary matrix spaces or with AI heuristics that rely on static analysis and prompt engineering~\cite{AutomaticRepairQuantum,repairGPT}. 
In contrast, Bloq assertions provide a data-driven mechanism for inferring and correcting faults based directly on the measured quantum state during testing and can be performed with minimal overhead.
This is because obtaining the fault-equivalent-matrix is a post-processing task which can be performed after Bloq assertion fault localization has been performed.
Thus, no further execution of the quantum circuit is required to perform classification or repair.

\subsection{Practical Considerations}
\label{sec:discussion_limitations}

Here, we discuss practical aspects of Bloq regarding its requirements and implementation on existing quantum hardware and quantum \glspl{sdk}.

\paragraph{\textbf{Quantum Hardware Support}}

Bloq is hardware-independent within the domain of gate-based quantum computing and can be applied to any quantum hardware platform supporting single-qubit and multi-qubit gate operations. 
This includes readily-available systems provided by vendors such as IBM, where superconducting qubit devices are accessible through the IBM Cloud platform with free or paid subscriptions, as well as trapped-ion systems available through Azure Quantum via providers such as IonQ and Quantinuum.

Since Bloq does not require mid-circuit measurements, its practical application is straightforward on current \gls{nisq} devices~\citep{ChenMidCircuit}. 
The fault localization procedure involves executing only segments of the circuit, terminated at the assertion point, followed by single-qubit Pauli measurements. 
These measurements require at most two additional single-qubit gates for measurement in the \(Y\) basis, typically implemented via a Hadamard and an \(S\) gate, making Bloq compatible with the native gate sets of current quantum hardware.

\paragraph{\textbf{Quantum Software Platforms}}

On the software side, Bloq can be implemented using any gate-based quantum \gls{sdk}~\citep{QiskitCite,cirq_developers_2023_10247207,joshua_combes_2019_3455848}. 
In this work, we have utilized IBM's Qiskit framework, and our implementation and experimental evaluations are available in our replication package~\cite{Bloq_replication}.
However, Bloq's principles are equally applicable to other \gls{sdk}s, such as Google's Cirq or Rigetti's pyQuil. 

The generation of Bloq assertions and their corresponding expected Pauli expectation values, derived from AutoBloq, are implemented in our evaluation as mathematical functions expressed through Python functions. 
These functions require as input a test case bitstring, the segment under test, and the specific qubit to generate the expected Bloch vector for assertion checking. 
This design ensures that Bloq can be easily integrated into existing quantum testing workflows, providing access to the circuit structure and the relevant input data.

\section{Related Work}
\label{sec:related_work}

This section reviews prior work on quantum software testing relevant to our approach. 
We organize the discussion into two subsections:
\begin{inparaenum}
    \item assertion-based quantum software testing
    and
    \item assertion placement and automation.
\end{inparaenum}

\subsection{Assertion-Based Quantum Software Testing}

Beyond Proq, prior assertion-based techniques include one that validates predicates uses 
statistical hypothesis testing on measurement results~\citep{Stat}, and one employs ancilla qubits and dynamic circuits to validate quantum state properties~\citep{QECA2020}.
\Citet{Stat} employs classical statistical testing on measurement results to validate predicates, while \citet{QECA2020} introduces auxiliary qubits to indirectly test the quantum state. 
However, \citet{QECA2020} utilize an exponentially-scaling oracle, making it impractical for larger circuits such as \gls{qft}. 
While \citet{QECA2020} provides experimental evaluation, it is limited to tiny circuits with ancilla qubits, unlike Bloq, which scales to larger programs such as Grover with 6 qubits and \gls{qft} with 10 qubits, both with substantial circuit depth.

Overall, Bloq complements and extends prior assertion-based approaches by providing an automated, scalable, and empirically-validated method for quantum program fault localization, tailored for current \gls{nisq} hardware limitations.

\subsection{Assertion Placement and Automation}

Assertion placement and automation are important challenges in quantum software testing, as noted by \citet{LiExploitingAssertions}. 

\citet{willeAssertionRefining} address assertion placement by relocating or inserting assertions to improve fault isolation. 
Their method defines commutation rules to move assertions earlier in the circuit, provided nearby gates do not affect their outcome.
This improves the usefulness of failing assertions without requiring manual adjustment.
Bloq complements this by generating assertion schemes automatically from high-level algorithm specifications using AutoBloq, avoiding the need for manual design. 
Since Bloq assertions use only simple unitaries to compute Bloch vector components, they cannot be further optimized through gate movement. 
However, \citet{willeAssertionRefining}'s relocation strategy could still be applied to the resulting scheme for improved placement.
While \citet{willeAssertionRefining} also add new assertions, these are derived by refining developer-supplied ones using interaction analysis.
In contrast, Bloq generates the full assertion scheme directly from the algorithm specification.

Another related approach is quAssert~\citep{quAssert}, which uses static analysis and sampling to insert assertions in structured subcircuits.
However, as noted by \citet{willeAssertionRefining}, quAssert assumes the input circuit is correct, limiting its ability to localize faults.
Bloq instead aligns assertions with the intended algorithm behavior, making it suitable for detecting deviations caused by faults.

\section{Conclusion and Future Work}
\label{sec:conclusion}

\glsresetall{}

This paper presented Bloq, a scalable and automated fault localization approach for quantum programs based on Bloch vector assertions.
Bloq leverages expectation value measurements of Pauli operators to enable low-overhead fault localization on current \gls{nisq} hardware, and introduces AutoBloq, a procedure for automatically generating assertion schemes from quantum algorithm specifications.

Our evaluation demonstrates that Bloq consistently outperforms the projective measurement-based approach Proq for Grover's algorithm, while achieving competitive performance for \gls{qft}.
Bloq's effectiveness increases with the number of qubits and circuit depth, and the approach maintains robustness under noise where Proq's performance degrades.
In addition, Bloq significantly reduces runtime and circuit depth overhead compared to Proq, making it particularly suitable for noisy quantum systems.
Finally, Bloq's automated assertion generation complements existing assertion engineering efforts, enabling the possibility of fully automated testing pipelines.

We outline three key directions for future work.
First, several features of Bloq can be explored for improvement. 
In particular, the current single-qubit assertions could be extended to multi-qubit assertions, allowing for even more efficient fault localization. 
Realizing this extension requires advancing AutoBloq to support multi-qubit assertion generation. 
A thorough analysis of target quantum algorithms must be performed to identify how the mathematical derivations of expected states can be systematically transformed into code. 
This involves performing the partial trace over all qubits except two or more, preserving the necessary qubit indices for multi-qubit density matrices.

Second, Bloq's fault classification and repair capabilities are well-positioned for future work. 
As outlined in our discussion, these features build on the measurement results already produced by Bloq assertions. 
Through post-processing techniques, fault classification could be performed by deriving a unitary transformation mapping the expected state to the measured state. 
Applying this transformation in the circuit and rerunning the program would allow Bloq to verify whether the assertion subsequently passes, enabling automated fault repair mechanisms.

Third, to enable Bloq for hybrid quantum-classical algorithms such as \gls{vqe} and \gls{qaoa}, the algorithm specification can be extended to account for intermediate measurement outcomes that influence the parameters of the quantum circuit in subsequent iterations.

\appendix

\section{A: Calculation for Grover}
\label{appendix:grover_calculation}

In this appendix, we provide the detailed derivation of the partial trace operation used in \cref{sec:autobloq_grover} to compute the reduced density matrix for a single qubit in Grover's algorithm.

To obtain the reduced density matrix for qubit $q$, we trace out all other qubits by performing a partial trace over the indices $j_k$ for $k \neq q$.

We begin with the diagonal term of the density matrix. 
The partial trace of the pure state $\ket{j_0 j_1 \cdots j_{n-1}} \bra{j_0 j_1 \cdots j_{n-1}}$ over all qubits except $q$ yields:

\begin{equation}
    \Tr_{j_k, k \neq q} \ket{j_0j_1\cdots j_{n-1}}\bra{j_0j_1\cdots j_{n-1}} = \ket{j_q}\bra{j_q}
\end{equation}

Next, we consider the off-diagonal terms. For instance, the partial trace of a term with differing bitstrings in the ket and bra takes the form:

\begin{equation}
    \Tr_{j_k, k \neq q} \sum_{x_0\cdots x_{n-1} \neq j_0\cdots j_{n-1}}\ket{x_0\cdots x_{n-1}}\bra{j_0\cdots j_{n-1}}
\end{equation}

The ket summation includes all possible basis states, excluding the solution state represented by the bra. 
However, due to the properties of the partial trace, only terms that are identical to the bra in all qubit indices except $q$ survive the trace. 
This means the only non-zero contribution comes from the basis state that differs from $\ket{j_0 \cdots j_{n-1}}$ only at position $q$. 

We denote the flipped bit at qubit $q$ as $\Tilde{j_q} \equiv j_q \oplus 1$, where $\oplus$ denotes bitwise XOR.

Accordingly, we obtain:

\begin{equation}
    \Tr_{j_k, k \neq q} \sum_{x_0\cdots x_{n-1} \neq j_0\cdots j_{n-1}}\ket{x_0\cdots x_{n-1}}\bra{j_0\cdots j_{n-1}} = \ket{\Tilde{j_q}}\bra{j_q}
\end{equation}

The second off-diagonal term follows by symmetry:

\begin{equation}
    \Tr_{j_k,k\neq q} \sum_{x_0\cdots x_{n-1} \neq j_0\cdots j_{n-1}} \ket{j_0\cdots j_{n-1}}\bra{x_0\cdots x_{n-1}}=\ket{j_q}\bra{\Tilde{j_q}}
\end{equation}

We now turn to the final term, which involves a sum over all basis states excluding the solution in both the ket and bra. 
This term is more involved and requires further manipulation. 
We begin by writing it explicitly as:

\begin{equation}
    \Tr_{j_k, k\neq q} \Big[ \Big( \sum_{x_0\cdots x_{n-1} \neq j_0\cdots j_{n-1}} \ket{x_0\cdots x_{n-1}} \Big)\Big( \sum_{y_0\cdots y_{n-1} \neq j_0\cdots j_{n-1}} \bra{y_0\cdots y_{n-1}} \Big) \Big]
\end{equation}

To simplify this expression, we rewrite the ket and bra sums by adding and subtracting the solution state $\ket{j_0 \cdots j_{n-1}}$, which allows us to express the sums as full sums over the entire basis minus the solution state:

\begin{equation}
    \Tr_{j_{k\neq q}}\Big[ \Big( \sum_{x_0x_1\cdots x_{n-1}} \ket{x_0x_1\cdots x_{n-1}} - \ket{j_0j_1\cdots j_{n-1}} \Big)
\end{equation}
\begin{equation}
 \Big( \sum_{y_0 y_1\cdots y_{n-1}} \bra{y_0y_1\cdots y_{n-1}} - \bra{j_0 j_1\cdots j_{n-1}} \Big)\Big]
\end{equation}

To proceed, we make use of the following identity:

\begin{equation}
    \Tr_{j_{k\neq q}} \Big( \sum_{y_0y_1\cdots y_{n-1}} \ket{k_0k_1\cdots k_{n-1}}\bra{y_0y_1\cdots y_{n-1}} \Big)
\end{equation}

\begin{equation}
    = \ket{k_q}\bra{0}+\ket{k_q}\bra{1}
\end{equation}

This identity reflects the fact that the partial trace over all qubits except $q$ produces a sum over bra states in the $q$-th register, preserving only the $q$-th qubit from the ket.

We now apply this to simplify the full expression:

\begin{equation}
    \Tr_{j_{k\neq q}} \Big[ \Big( \sum_{x_0x_1\cdots x_{n-1}} \ket{x_0x_1\cdots x_{n-1}} \Big)
\end{equation}
\begin{equation}
    \Big( \sum_{y_0y_1\cdots y_{n-1}}\bra{y_0y_1\cdots y_{n-1}} \Big) \Big]
\end{equation}

Since the trace is a linear operation, we can move the first summation outside the trace:

\begin{equation}
    \sum_{x_0x_1\cdots x_{n-1}} \Tr_{j_{k\neq q}} \Big( \sum_{y_0y_1\cdots y_{n-1}}\ket{x_0x_1\cdots x_{n-1}}\bra{y_0y_1\cdots y_{n-1}} \Big)
\end{equation}

Applying the identity from above, we obtain:

\begin{equation}
    =\sum_{x_0x_1\cdots x_{n-1}} \Big( \ket{x_q}\bra{0} + \ket{x_q}\bra{1} \Big)
\end{equation}

Expanding this into explicit sums over each qubit index:

\begin{equation}
    = \sum_{x_0=0}^1 \sum_{x_1=0}^1 \cdots \sum_{x_q=0}^1 \cdots \sum_{x_{n-1}=0}^1 \Big( \ket{x_q}\bra{0}+\ket{x_q}\bra{1} \Big)
\end{equation}

Since all qubit indices except $q$ are traced over, their corresponding sums contribute a multiplicative factor of $2$ each. 
There are $n - 1$ such qubits, resulting in a total of $2^{n - 1} = N / 2$ terms. Thus, we obtain:

\begin{equation}
    = \frac{N}{2}\Big(\mathbbm{1} + X\Big)
\end{equation}

The next two terms are:

\begin{equation}
    \Tr_{j_{k\neq q}}\Big(\sum_{x_0x_1\cdots x_{n-1}}\ket{x_0x_1\cdots x_{n-1}}\bra{j_0j_1\cdots j_{n-1}}\Big)=\ket{0}\bra{j_q} + \ket{1}\bra{j_q}
\end{equation}

\begin{equation}
    \Tr_{j_{k\neq q}}\Big(\sum_{y_0y_1\cdots y_{n-1}}\ket{j_0j_1\cdots j_{n-1}}\bra{y_0y_1\cdots y_{n-1}}\Big)=\ket{j_q}\bra{0}+\ket{j_q}\bra{1}
\end{equation}

The final term is straightforward:

\begin{equation}
    \Tr_{j_{k\neq q}}\Big( \ket{j_0j_1\cdots j_{n-1}}\bra{j_0j_1\cdots j_{n-1}} \Big) = \ket{j_q}\bra{j_q}
\end{equation}

Combining all contributions, we obtain the result for the first term in the density matrix expansion:

\begin{equation}
\Tr_{j_{k\neq q}} \Big( \ket{\alpha}\bra{\alpha} \Big) = \frac{\cos^2\phi_k}{N-1} \Big[ \frac{N}{2}\Big( \mathbbm{1}+X \Big) + \mathbbm{1} + X + (-1)^{j_q} Z\Big]
\end{equation}

Thus, the total reduced density matrix for the $k$-th Grover iteration and for qubit $q$ is given by:

\begin{align}
\rho_{k,q} &= \frac{\cos^2\phi_k}{N-1} \Big[ (\frac{1}{2}(N-1)\mathbbm{1}+(\frac{N}{2}-1)X-\frac{1}{2}(-1)^{j_q}Z\Big] \\
&+ \frac{\cos\phi_k \sin\phi_k}{\sqrt{N-1}} X + \sin^2\phi_k \frac{1}{2}\Big( \mathbbm{1} + (-1)^{j_q}Z \Big) 
\end{align}

Finally, this expression simplifies to the compact form presented in \cref{eq:density_matrix_expected_grover} in the main text.


%



\bibliographystyle{plainnat}
\bibliography{biblio}
%

\end{document}